# X-RAY TEMPERATURES FOR THE EMSS HIGH REDSHIFT CLUSTER SAMPLE: CONSTRAINTS ON COSMOLOGY AND THE DARK ENERGY EQUATION OF STATE

J. Patrick Henry
Institute for Astronomy, University of Hawaii, 2680 Woodlawn Drive, Honolulu, HI 96822;
henry@ifa.hawaii.edu

## ABSTRACT

We measure the X-ray temperature (and luminosity) with ASCA of all but one cluster in the *Einstein* Extended Medium Sensitivity Survey (EMSS) high redshift ($z \geq 0.3$) sample. We compare these data to a complete sample of low redshift clusters that also has temperature measurements, thereby providing cosmological constraints. Improvements over our previous work include: (1) An enlarged high redshift sample; (2) Temperatures for the low redshift comparison sample come from the same instrument as the high redshift sample; (3) The elimination of three EMSS clusters with the same redshift as the target (i.e. not truly serendipitous) and a fourth with an ASCA flux well below the completeness limit; (4) Using a theoretical cluster mass function that more closely matches N-body simulations (the Sheth – Torman function); (5) Using a cold dark matter power spectrum instead of a power law; (6) Using a general cosmology with arbitrary matter density and cosmological constant; (7) Using a cosmology that generalizes the cosmological constant to quintessence; (8) Including the effects of temperature measurement errors and scatter in the cluster luminosity - temperature relation; (9) Marginalizing over the poorly known normalization of the mass - temperature relation.

We find an allowed band in the $\Omega_{m0}$ - $\Omega_{\Lambda0}$ plane of different orientation to the band of constraints provided by the supernovae Ia Hubble diagram and the cosmic microwave background fluctuations. All three bands intersect at the same place: $\Omega_{m0} \approx 0.3$, $\Omega_{\Lambda0} \approx 0.7$. We measure the quintessence equation of state parameter to be $w = -(0.42 \pm 0.21)$ (68% confidence for 1 interesting parameter), consistent with previously determined upper limits. We measure the normalization of the mass fluctuation power spectrum to be $\sigma_8 = 0.66 \pm 0.16$ (68% confidence for 3 interesting parameters). Systematic errors are larger than the statistical errors only for $\sigma_8$ with our sample, thus the errors for it depends on the details of the marginalization over the temperature – mass normalization.

## 1. INTRODUCTION

The much-anticipated first results from WMAP herald the era of precision cosmology (Bennett et al., 2003; Spergel et al., 2003). Spergel et al. (2003) conclude their paper by comparing the current situation in cosmology with that in particle physics thirty years ago when particle physicists converged on a standard model. Cosmology now has such a standard model, with a small number of relatively precisely determined parameters that can fit a wide range of data but does not answer fundamental questions. Among these questions are the nature of the dark matter and dark energy. Particle physics did not end with the elucidation of its standard model and neither will cosmology. The task now confronting cosmologists is to rigorously test the standard model to find its failures thereby searching for the new physics that answers the fundamental questions.

Measurements of the properties of clusters of galaxies and the evolution of these properties can provide such tests because these measurements yield strong cosmological constraints that are independent of those used to determine the parameters of the standard model. There are three principal cluster measurements that have been used for this purpose. The first is the power spectrum of the three-dimensional distribution of clusters (Schuecker et al., 2003a and Schuecker et al., 2003b are recent examples). The second is the baryon fraction and its evolution (Allen et al., 2003a; Allen, Schmidt & Bridle, 2003b; Ettori, Tozzi & Rosati, 2003; Lin, Mohr & Stanford, 2003 are recent examples). The third is the cluster number density and its evolution (Ikebe et al., 2002; Reiprich & Böhringer, 2002; Rosati, Borgani & Norman 2002; Bahcall et al., 2003; Pierpaoli et al., 2003; Shimizu et al., 2003; Viana et al., 2003; Vikhlinin et al., 2003a are recent examples). Current samples tightly constrain the normalization of the matter fluctuation power spectrum parameterized by $\sigma_8$, the rms mass fluctuation in spheres of radius $8h^{-1}$ Mpc, and $\Omega_{m0}$, the present ratio of the mass density of the universe to the critical density. Here h is the present value of the Hubble parameter in units of 100 km $s^{-1}$ $Mpc^{-1}$. Less tightly constrained are the cosmological constant or dark energy, $\Omega_{\Lambda 0}$, and its equation of state parameter w. Nevertheless, the constraints on these two parameters are complementary to those from observations of supernovae, the microwave background and gravitational lensing, particularly for $\Omega_{\Lambda 0}$.

We have been using observations of the cluster number density as a function of temperatures to constrain cosmology in a series of papers of increasing sophistication. Henry & Arnaud (1991, hereafter HA) analyzed a local sample and assumed $\Omega_{m0} = 1$. Henry (1997) and Henry (2000, hereafter H00) analyzed both local and distant samples, thereby measuring evolution, and assumed either $\Omega_{m0} \leq 1$ and $\Omega_{\Lambda 0} = 0$ or $\Omega_{m0} + \Omega_{\Lambda 0} = 1$. The present paper builds on H00. We extend the high-redshift sample to the entire EMSS $z \geq 0.3$ sample, which is the largest sample at these redshifts with (nearly) complete temperature information. We also use measurements of the temperatures of all the local clusters with the same instrument used for the distant objects. We use an analytic mass function that provides a closer match to numerical N-body simulations and we included the effects of temperature errors and scatter in the luminosity-temperature relation in our analysis. Perhaps the most important improvement is the extension of the underlying theory to a wider region of parameter space. In particular, we do not assume anything for the values of $\Omega_{m0}$, $\Omega_{\Lambda 0}$ and w (within reasonable bounds).

## 2. OBSERVATIONS

We want to measure the evolution of cluster temperatures. Since clusters are not constant temperature baths, we must construct the cluster temperature function in order to compare the populations at different epochs. We present the data as integral temperature functions n(>kT), i.e. the number of clusters per unit volume hotter than a given temperature. This form has the advantage that it is nonparametric and unbinned. Its main disadvantage is that all points except the hottest are correlated. That means that not only can an individual data point move within its error range, but also all cooler data points move along with it. This disadvantage is not too serious when the temperature errors are small, because the very steep dependence of n(>kT) on temperature insures that data even a factor of two different in temperature are almost uncorrelated. Some of our high redshift cluster temperature errors are not small and this makes the graphical comparison of data and model more difficult.

We estimate the integral temperature function and its error from

$$n(>kT) = \sum_i \frac{1}{V_{sea,i}} \quad (1)$$

$$\sigma^2[n(>kT)] = \sum_i \frac{1}{V^2_{sea,i}} \quad (2)$$

where the sum is over all clusters with $kT_i > kT$ (*not* with $kT_i \geq kT$) and $V_{sea,i}$ is the search volume in which the $i^{th}$ cluster could have been detected.

## 2.1 *Low-Redshift Sample*

The low-redshift sample remains the 25 clusters from HA91. We impose the additional restriction kT > 3 keV because the heating of the intracluster medium may not be dominated by gravity below that temperature. This restriction removes the Virgo cluster from the HA91 sample leaving 24 objects. The average redshift is 0.051. These clusters were selected originally from the early all-sky surveys made with nonimaging detectors, augmented by pointed observations with other nonimaging detectors but with greatly reduced fields of view. All but the Coma cluster are unresolved with these instruments, so accurate total fluxes are available for this sample. Clusters are of course diffuse sources so obtaining total fluxes for them from images is not a simple task. It is for this reason that we retain this particular low-redshift sample.

Larger low-redshift samples are available. For example, Ikebe et al. (2002) have presented a sample of 61 clusters, all but 4 of which have measured temperatures. Our entire sample is contained in theirs, but most of their sample comes from ROSAT imaging and some extrapolation is required to obtain the total flux. In Section 5, after we describe our methods, we compare the constraints on the two best-determined cosmological parameters provided both by our low redshift sample and the Ikebe et al. sample. This comparison shows that the two samples provide almost identical constraints and therefore the choice of low redshift samples does not have a large impact on the final cosmological constraints.

The selection criteria for the low-redshift sample are that the total flux in the 2-10 keV band is greater than 3 x $10^{-11}$ ergs $cm^{-2}$ $s^{-1}$, the redshift is less than 0.2, the temperature is greater than 3 keV and the absolute galactic latitude is greater than $20^o$ (minus a small zone of exclusion around the Magellanic Clouds), yielding a solid angle surveyed of 8.23 sr. An improvement over H00 is that all clusters in this sample now have integrated temperature measurements determined from ASCA data (White 2000). We also used ASCA to measure the temperatures of the high redshift sample.

Many clusters have cooling centers resulting from the high gas density there. It is likely that the gas temperatures uncontaminated by these centers are a better proxy for the cluster mass, the fundamental variable in the theory (see §4). It is possible at low redshift to remove the contribution of the cooling center, either by excluding it spatially (Markevitch, 1998; Fukazawa et al. 1998) or by fitting the integrated spectrum with a sufficiently complex model (Allen & Fabian 1998; Ikebe et al. 2002). It has only become possible to do one or the other of these at high redshift with the advent of Chandra and XMM-Newton, but no complete high-redshift sample yet has such observations. Consequently, in order to compare the low- and high- redshift samples, we use the integrated temperature for the low redshift sample as well. This requirement precludes our use of the Ikebe et al. temperatures, which are only given for cool center corrections. We give the properties of the low-redshift sample in Table 1, including the search volumes determined using the method we describe in §3.1.

## 2.2 *High-Redshift Sample*

The high-redshift sample comprises the clusters with $z \geq 0.3$ from the *Einstein* Extended Medium Sensitivity Survey (EMSS) (Gioia et al. 1990; Henry et al. 1992, hereafter H92; Gioia & Luppino 1994, hereafter GL94). Sources in the EMSS are detected serendipitously in *Einstein* Observatory Imaging Proportional Counter (IPC) fields targeted at unrelated objects. Each source in our sample had a sufficient number of counts in the EMSS 2.4′ x 2.4′ detect cell that its 0.3 – 3.5 keV band flux was measured at > $4\sigma$ significance and the value of that flux was $\geq$ 1.33 x $10^{-13}$ ergs $cm^{-2}$ $s^{-1}$. An additional selection was that $\delta(1950) \geq -40^{o}$. These criteria yield the integral sky coverage or solid angle searched as a function of detect cell flux given in Table 3 of H92.

There are 26 EMSS clusters with z ≥ 0.3 in GL94. We eliminate MS1209.0 for the reason described in the note added to GL94 (see also Rector et al. 2000), and MS 1333.3 and MS1610.4 because they are X-ray point sources. Further we exclude MS0147.8 because its flux measured in our ASCA observation is substantially below the above flux limit. All but 3 of the remaining 22 sources have ROSAT HRI data showing they are extended, i.e. *bona fide* clusters. Those objects without HRI data are MS0418.3, MS0821.5 and MS1532.5.

We then examined the targets in whose fields the remaining 22 sources were found to determine whether the cluster had the same redshift. If the redshifts are the same, then the EMSS source is not serendipitous and should also be removed from our sample. This procedure is not as straightforward as one might think. Consider MS0015.9 = Cl0016+16, which was found in the field of a cluster candidate Cl0014+15. Subsequent Keck multiobject spectral observations of the candidate by D. Koo showed no concentration of redshifts, i.e. the target cluster does not exist. It is hard to imagine that MS0015.9 is associated with an object that does not exist and so we retain it. On the other hand, the overdensity of galaxies that produced the candidate cluster Cl0014+15 likely occurred because of a wall-like structure in this region that includes Cl0016+16 (Connolly et al. 1996). So is MS0015.9 truly serendipitous? Fortunately, no other case is as ambiguous. We did find two targets, involving three additional EMSS clusters that have matching redshifts. The two clusters MS0302.5 (z = 0.4249) and MS0302.7 (z = 0.4246) form a supercluster with the target Cl0303+17 (z = 0.418, Kaiser et al. 1998). MS1512.4 (z = 0.3726) was found in the field whose target was 4C37.43 (z = 0.3707). It is interesting that MS1512.4 was the source most discrepant with respect to the temperature selection function in Figure 5 of H00. These three clusters have been removed from the sample, but we give the results of their ASCA observations in Table 2.

Now we discuss the case of Cl0152.7, originally discovered by the WARPS survey (Ebeling et al. 2000). This cluster is at z = 0.8325. It was not in the EMSS because of a detail in the way that survey measured positions and flux significances. Positions were determined in the IPC hard band because that band had the tightest point spread function. Flux significances were determined in the total band, because that band was more sensitive, centered at the hard band position. In the case of Cl0152.7, the hard position yields a flux significance just below $4\sigma$, but the total band position yields a significance just above that level. This source should have been in the EMSS, see Ebeling et al. (2000) for a detailed discussion. However a Chandra observation revealed that the IPC source is confused (Maughan et al., 2003). There are two clusters at the same redshift, neither meeting the EMSS criteria for this field. Thus in the end we exclude Cl0152.7 from our complete sample, leaving the 19 clusters at the top of Table 2. Their average redshift is 0.429.

We have determined temperatures, fluxes and luminosities for the EMSS high redshift sample from ASCA observations using the procedure described in H00. We give the results in Table 2. Briefly, spectra were extracted from 6.125′ and 2.5′ radius regions for the Gas Imaging Spectrometer (GIS) and Solid-State Imaging Spectrometer (SIS) respectively. Background was acquired from adjacent regions of the same exposure. Counts were grouped until each bin contained at least 20 counts. We fit the four spectra, two from each GIS and SIS, simultaneously. The adjustable parameters were the GIS and SIS normalizations, temperature, abundance and SIS hydrogen column density. Fixed parameters were the redshift and GIS hydrogen column density (set to the 21 cm value). The total flux and luminosity were determined from the GIS normalization assuming the clusters are point sources, a good approximation for clusters at these redshifts observed by the GIS on ASCA.

Distant clusters are unresolved with ASCA observations. Hence, their total flux may be measured relatively straightforwardly with a procedure that is independent of the object's redshift, making the high redshift fluxes more comparable to those at low redshift. However because the EMSS fluxes are measured in the 2.4′ x 2.4′ detect cell, some method must be used to account for the flux outside the detect cell to determine the total cluster flux. H92 describe this method in detail. We summarize it here because we need it to derive the temperature selection function and we can assess its validity with our ASCA data. We assume all clusters have the same intrinsic surface brightness, $I(r) = I_0[1 + (r/a)^2]^{-3\beta+0.5}$ with $\beta = 2/3$ and core radius $a = 0.125\ h^{-1}$ Mpc. The standard EMSS procedure is to integrate this surface brightness over the detect cell to obtain the detect cell flux, $F_{det}$, and out to infinity to obtain the total flux, $F_{tot}$. We define $f = F_{det}/F_{tot}$ and the size correction $SC = f^{-1}$. No cluster extends to infinity, so it would have been better to integrate out to the virial radius to obtain $F_{tot}$, but we want to evaluate the standard procedure first. We will discuss this issue momentarily. With this definition f is

$$f\left(\frac{\theta_D}{\theta_0}\right) = \frac{2}{\pi}\sin^{-1}\left(\frac{\theta_D^2/\theta_0^2}{\theta_D^2/\theta_0^2+1}\right) \qquad (3)$$

where $\theta_D$ is the angular half-size of the detect cell, 1.2′, and $\theta_0$ is the angular size of the cluster core radius.

We show in Figure 1 the ratio $F_{tot}$(0.3,3.5;ASCA) / $F_{tot}$(0.3,3,5;EMSS) = $F_{tot}$(0.3,3.5;ASCA) f /$F_{det}$(0.3,3.5;EMSS) ≡ $F_{det}$(0.3,3.5;ASCA) / $F_{det}$(0.3,3,5;EMSS) where the total or detect cell fluxes in the 0.3 – 3.5 keV band are from the ASCA observations here or from the EMSS. The weighted average of this ratio is 1.15 ± 0.05. If we include only the 5 sources not contaminated by HRI point sources within the 6.125′ radius region over which the ASCA counts were accumulated, then the ratio is 0.99 ± 0.08. Ellis & Jones (2002) have performed the same analysis, but comparing the EMSS fluxes to those measured with the ROSAT PSPC for 9 $z \geq 0.3$ clusters. They find $F_{tot}$(0.3,3.5;ROSAT) / $F_{tot}$(0.3,3,5;EMSS) = 1.18 ± 0.08 if the ROSAT counts are extrapolated to infinity or 1.05 ± 0.07 if extrapolated to the virial radius. Clearly all these ratios are consistent with each other and show empirically that the EMSS correction from detect cell to total fluxes is correct on average at the ~10% level. The average statistical error on the EMSS flux for the clusters in Table 2 is 18%. We therefore adopt the standard EMSS size correction because it is empirically verified to be correct at a level equal to or less than the EMSS statistical errors.

This is not to say that this correction is perfect. No procedure that uses a single X-ray surface brightness distribution for all clusters can be since all clusters are not the same. In particular, Ellis & Jones (2002) find that there is a trend for the above ratio to be larger for clusters with larger core radii. The standard procedure also did not include the Einstein IPC point spread function, which is 1.5′ (FWHM) compared to the 2.4′ full width of the detect cell. However as H00 and Ellis & Jones (2002) note, including the PSF and only extrapolating the flux to the virial radius fortuitously nearly cancel.

Is our high-redshift EMSS sample statistically complete? This question has been raised repeatedly since the original EMSS publications, prompted at least in part by an interpretation of the lack of high redshift clusters as incompleteness rather than evolution. The most recent example is by Lewis, Ellingson & Stocke (2002) who claim that the EMSS is 77 ± 6% complete, having "systematically missed clusters of low surface brightness." It is difficult to prove or disprove this assertion, since it amounts to only 4 missing objects in our sample of 19. While the EMSS was the only sample of high redshift X-ray selected clusters available for some time, there are now several such samples constructed from the ROSAT database. The consensus of this recent work is that clusters do evolve, about as the EMSS claimed (see Rosati et al., 2002 for a review). As an example, consider the Massive Cluster Survey (MACS, Ebeling, Edge & Henry, 2001). This high redshift sample is the largest to date. MACS finds an integral luminosity function $n(>10^{45}\ \mathrm{erg\ s^{-1}}) = 3.53^{+0.87}_{-0.64} \times 10^{-9}\ \mathrm{Mpc^{-3}}$ averaged over the redshift interval 0.4 – 0.5. The EMSS value is 3.26±1.88 x $10^{-9}$ $Mpc^{-3}$ averaged over the redshift interval 0.3 – 0.6 (average z = 0.466) and both for h = 0.5, $\Omega_{m0}$ = 1, $\Omega_{\Lambda 0}$ = 0. Although this agreement does not prove that the EMSS is complete, such agreement of independent samples is the usual criterion for correctness in astronomy.

## 3.0 TEMPERATURE SELECTION FUNCTION AND MAXIMUN LIKELIHOOD FITTING

### 3.1 *From Flux Selection to Temperature and Redshift Selection*

We describe how to determine the temperature selection function in this section. The method is more general than that given in H00 and may be used for any X-ray flux limited survey. All X-ray selected clusters so far come from flux-limited, not temperature-limited, surveys. That is the solid angle surveyed is known as a function of flux in some band and in some detect cell. The detect cell flux may be converted to luminosity and redshift in the usual way yielding the solid angle surveyed in which a cluster with these properties could have been detected (the integral solid angle in the terminology of H00).

$$\Omega(F_{det}(E_1,E_2)) = \Omega(L(bol),z) = \Omega\left(\frac{L(bol)\;BF(E_1,E_2,kT(L))}{4\pi\;D_L^2(\Omega_{m0},\Omega_{\Lambda 0},w,z)\;SC(z)\;k(E_1,E_2,z,kT(L))}\right) \quad (4)$$

Here L(bol) is the bolometric luminosity, BF is the band fraction that gives the fraction of the bolometric luminosity that is in the energy band $E_1$ to $E_2$, $D_L$ is the luminosity distance, SC is the size correction, the ratio of the total flux from the cluster to the flux in the detect cell, and k is the k correction. We use the temperature from cluster temperature – luminosity relation when determining the band fraction and k correction. We assume a constant metric core radius when determining the SC but convert it to an angular size using the cosmology being fitted. While the flux selection function is cosmology independent, the selection function of luminosity and redshift is not.

The cluster temperature – luminosity (and luminosity – temperature) relation exhibits substantial scatter (e.g. Novicki, Sornig, and Henry, 2002). If there were no scatter, then we could simply replace L(bol) in the above relation with a unique temperature, thereby obtaining a selection function of temperature and redshift. Because the band fraction and k correction are weak functions of temperature for the small range of temperatures considered here, the scatter in the temperature – luminosity relation produces almost no scatter for them. Thus to obtain Ω(kT,z) we must average over the possible Ω(L(bol),z)s at each temperature, weighted by the probability of obtaining them. The scatter in L(bol) at constant kT is Gaussian in log(L(bol)) with dispersion σ(logL) (Novicki et al. 2002), hence

$$\Omega(kT,z) = \int d(\log L)\;\Omega(L,z)\;\frac{\exp\left\{-\left[\log(A(kT)^{\alpha}D_L^2(\Omega_{m0},\Omega_{\Lambda 0},w,z)/D_L^2(1,0,-1,z)) - \log L\right]^2/2\sigma^2(\log L)\right\}}{\sqrt{2\pi\sigma^2(\log L)}} \quad (5)$$

where we have assumed that the luminosity – temperature relation is $L(bol) = A(kT)^{\alpha}\,[D_L(\Omega_{m0}, \Omega_{\Lambda 0}, w, z)/D_L(1, 0, -1, z)]^2$. The luminosity distance dependence converts from the $\Omega_{m0}$ =1, $\Omega_{\Lambda 0}$ =0 cosmology in which the A and α are determined and also makes the temperature selection function cosmology dependent.

The specific relations we use for the luminosity – temperature relation and its inverse come from White, Jones & Forman (1997): $L_{44}(bol) = 4.78\times10^{-2}\,(kT)^{2.98}\,[D_L(\Omega_{m0}, \Omega_{\Lambda 0}, w, z)/D_L(1, 0, -1, z)]^2$ and $kT = 2.774[L_{44}(bol)]^{0.336}\,[D_L(1, 0, -1, z)/D_L(\Omega_{m0}, \Omega_{\Lambda 0}, w, z)]^{0.671}$ where $L_{44}$ is the luminosity in units of $10^{44}$ erg $s^{-1}$. We have modified the coefficient and exponents of the latter relation from that found by White et al. (1997) by ~1% in order to make the two exact inverses of each other. The specific value of the dispersion comes from Novicki et al. (2002): σ(logL(bol)) = 0.285. We assume that the luminosity – temperature relation does not evolve, as the data in Novicki et al. indicate. Vikhlinin et al. (2002) show that this relation does evolve when considering temperatures and luminosities corrected for cooling centers. They find $L \sim (1 + z)^{0.6}$. As discussed in Section 2.1 it is not possible to make these corrections for our distant clusters, so this result is not directly applicable.

The search volume can be calculated two ways.

$$V_{sea}(L) = \int_{z\,min}^{z\,max} dz\, \Omega(L,z) \frac{d^2V}{d\Omega dz} \text{ or } V_{sea}(kT) = \int_{z\,min}^{z\,max} dz\, \Omega(kT,z) \frac{d^2V}{d\Omega dz} \qquad (6)$$

where the luminosity and temperature of the cluster searched for are L and kT respectively. Correspondingly, the temperature function may be constructed using either $1/V_{sea}(L)$ or $1/V_{sea}(kT)$. Ikebe et al. (2002) and Markevitch (1998) used Monte Carlo simulations to show both estimators are consistent and unbiased. They reproduce the true temperature function in the limit of an infinite number of clusters in a sample. Further the mean temperature function obtained from a number of independent samples with a finite number of clusters is also equal to the true temperature function in the limit of an infinite number of samples. However, the $1/V_{sea}(kT)$ method gives a smaller variance and is thus more efficient. We use it here and show the resulting temperature functions in Figure 2. We may give an indication of the magnitude of the difference between the two search volumes by considering the Euclidean limit. In that case $V_{sea}(kT)/V_{sea}(L) \approx 1 + 2\sigma^2(\log L)$ or a 16% effect for the σ(logL(bol)) used here.

### 3.2 *Maximum Likelihood Fitting*

We generalize the maximum likelihood method of Marshall et al. (1983) to the case where there are errors on the measured temperatures σ(kT) for N clusters in the sample; σ (kT) may include a root mean square sum contribution from the scatter in the mass – temperature relation if desired.

$$S = -2\sum_{i=1}^{N} \ln\left[\int dkT\, n(\Omega_{m0},\Omega_{\Lambda 0},\sigma_8,w,z_i,kT)\, \frac{\exp\left[-\left(kT_i - kT\right)^2 / 2\sigma_i^2(kT)\right]}{\sqrt{2\pi\sigma_i^2(kT)}}\, \Omega(kT,z_i) \frac{d^2V}{d\Omega dz}\right]$$

$$+\,2 \int_{kT\,min}^{kT\,max} dkT \int_{z\,min}^{z\,max} dz\, n(\Omega_{m0},\Omega_{\Lambda 0},\sigma_8,w,z,kT)\Omega(kT,z) \frac{d^2V}{dz d\Omega} \qquad (7)$$

This expression reduces to equation (6) on H00 in the limit that the temperature errors are very small and the selection function is independent of cosmology. The integral of the Gaussian distribution is absent from the second term of S because it is possible to perform it analytically. The best estimates of the model parameters are then obtained by minimizing S. Confidence regions for the best estimates are obtained by noting that S is distributed as $\chi^2$ with the number of degrees of freedom equal to the number of interesting parameters. The values of $z_{min}$, $z_{maz}$, $kT_{min}$ and $kT_{max}$ are 0.00 – 0.20, 3 – 10 keV and 0.30 – 0.85, 4 – 12 keV for the low and high redshift samples respectively.

## 4.0 THEORETICAL TEMPERATURE FUNCTION

We give a general summary of the theoretical temperature function and its evolution in this section. The details of various functions and parameters for the models under consideration are in the Appendix. The notation here is somewhat more compact than H00 and so hopefully is more transparent. We start with the commoving mass function of collapsed objects.

$$n(\Omega_{m0},\Omega_{\Lambda 0},\sigma_8,w,z,M)=\frac{\rho_{b0}}{M}\frac{d\nu}{dM}f(\nu) \quad (8)$$

Here $\rho_{b0}$ is the present value of the background mass density and

$$\nu(\Omega_{m0},\Omega_{\Lambda 0},\sigma_8,w,z,M)=\frac{\delta_c(\Omega_{m0},\Omega_{\Lambda 0},w,z)}{\sigma(\Omega_{m0},\Omega_{\Lambda 0},\sigma_8,w,z,M)} \quad (9)$$

$$\sigma(\Omega_{m0},\Omega_{\Lambda 0},\sigma_8,w,z,M)=\sigma_8\left[\frac{1+2.208m^p-0.7668m^{2p}+0.7949m^{3p}}{1+2.208m_8^p-0.7668m_8^{2p}+0.7949m_8^{3p}}\right]^{-2/9p}\frac{D(\Omega_{m0},\Omega_{\Lambda 0},w,z)}{D(\Omega_{m0},\Omega_{\Lambda 0},w,0)} \quad (10)$$

In these expressions $\delta_c$ is the mass density fluctuation required for a spherical perturbation to collapse at redshift z and D is the growth factor of perturbations while they are evolving linearly. The rms mass fluctuations in spheres containing mass M are $\sigma$. The mass fluctuation power spectrum, P(k) determines $\sigma$. The quantity $\sigma_8$ is related to the normalization of the power spectrum by $\sigma_8 \approx [P(0.172\ h\ Mpc^{-1})/3879\ h^{-3}\ Mpc^3]^{1/2}$ (Peacock 1999, equations 16.13 and 16.132). We assume that the shape of P(k) comes from cold dark matter. Specifically we use the approximation from Kitayama & Suto (1996) in equation (10). With this approximation p = 0.0873, m = 1000 $hM_{15}\Gamma^2h$, $m_8$ = 597.2 $\Omega_{m0}\Gamma^2h$ with $\Gamma=\Omega_{m0}h(T_0/2.7K)^{-2}\exp[-\Omega_{b0}(1+\sqrt{2h}/\Omega_{m0})]$ and $M_{15}$ the cluster mass in units of $10^{15}$ M⊙. Adopting a microwave background temperature of $T_0$ = 2.726 ± 0.010 (Mather et al. 1994), a baryon density of $\Omega_{b0}$ = 0.044 ± 0.004 and h = 0.71 ± 0.04 from the standard cosmological model (Spergel et al., 2003) yields $m=344.5\Omega_{m0}^2\exp[-0.088(1+1.192/\Omega_{m0})]hM_{15}$ and $m_8=205.7\Omega_{m0}^3\exp[-0.088(1+1.192/\Omega_{m0})]$. The errors on $T_0$, $\Omega_{b0}$, and h yield a ± 15% error on the coefficients of m and $m_8$, dominated by the errors on h. It might be worthwhile marginalizing over h if its error range induced large changes in the theoretical temperature function. However there is no change in $\nu$ from equation (10) because of the same dependence in both numerator and denominator. The change on the temperature function normalization from propagating the errors on these three quantities in equation (16) below is only ±1.5%, much less than the Poisson error from our sample of 43 objects. Expressed another way, the largest changes in the temperature function are caused by changes in the fitting parameters, not those that we assume as given. Thus we do not marginalize over the three parameters whose values come from the standard model.

Two often-used mass functions are

$$f(\nu)=\sqrt{\frac{2}{\pi}}e^{-0.5\nu^2} \quad (11)$$

$$f(\nu)=0.2162(1+1.1096\nu^{-0.6})e^{-0.3535\nu^2} \quad (12)$$

The first expression above is from Press & Schechter (1974). The second comes from Sheth & Torman (1999, ST). The main difference for our purposes is ST has more high mass clusters. It provides a better match to N body simulations (e.g. Jenkins et al. 2001). Jenkins et al. (2001) give a fitting function that describes their simulations slightly better than ST, but presumably different simulations will have different fitting functions. For this reason we use ST as our preferred mass function.

A mass – temperature relation is needed in order to convert the mass function to a temperature function.

$$kT = \frac{7.98\text{keV}}{\beta_{TM}}\left[\frac{\Omega_{m0}\Delta(\Omega_{m0},\Omega_{\Lambda 0},w,z_v)}{18\pi^2}\right]^{1/3}(hM_{15})^{2/3}(1+z_v) \quad (13)$$

Here $\Delta$ is the ratio of the cluster's average density to that of the background mass density, $z_v$ is the redshift at which the cluster virialized and $\beta_{TM}$ is a modification factor accounting for departures from virial equilibrium. Note that some authors use $\Delta_c$, the density ratio with respect to the critical density; the relation between the two density ratios is $\Delta_c = \Omega_m\Delta$.

There has been considerable discussion of the reliability of this mass - temperature relation. Vikhlinin et al. (2002) find that the gas mass - temperature relation does not evolve as equation (13). However equation (13) is for the total mass, not the gas mass, so this finding is not directly relevant here. It would be relevant if the gas mass fraction of the total mass were the same for all clusters at all times. This assumption is known not to be true at low redshift. The baryon fraction, nearly the same as the gas fraction, is lower for clusters with temperatures less than 4 keV (Lin et al., 2003). The value of $\beta_{TM}$ is also uncertain. We collect various determinations of it in Table 3, which is an update to Table 5 of H00. We only include simulations that correlate the emission-weighted temperature, because that is what is observed, against the virial mass. It appears that there is reasonable agreement among the simulations except that Bryan & Norman (1998, reference 4) seem somewhat high. Excluding reference 4, the average of the remaining simulations is $\beta_{TM} = 1.03 \pm 0.04$ (68% confidence error on the mean). The last five rows of Table 3 are empirical determinations, a considerable increase over H00. Allen et al. (2002, reference 13) did not calculate a $\beta_{TM}$. We determined a value by combining the mass of all objects in their sample with ASCA temperatures from Horner (2001). We excluded A141, A209 and A1351 because they do not have ASCA temperatures. The error-weighted average of the last three rows in Table 3 is 0.72 ± 0.04 (68% confidence error on the mean). We have excluded from the average the results from Hjorth et al. (1998) because they require a rather large extrapolation to the virial radius and from Horner et al. (1999) because all but one object in their sample is contained in the larger sample of Finoguenov et al. (2001). We will compare results using the average empirical $\beta_{TM}$ with the average $\beta_{TM}$ from the simulations in order to show the effects of this uncertainty when the effect are large. However, as recommended by Pierpaoli et al. (2003), our baseline results are obtained by marginalizing this uncertain parameter, which we do over the range $\beta_{TM} = 0.7 – 1.0$ assumed uniformly distributed.

From the discussion in §3.1 and here we see that there are some uncertainties in two key ingredients of our analysis: the luminosity – temperature and mass – temperature relations. We have adopted the simplest form for these relations applicable to our case that is not definitely ruled out by observations. However we need to be mindful that these relations may need revision when better data are available. Indeed, deviations from cluster scaling relations may provide cosmological information as well (Verde, Haiman & Spergel 2002).

We derive the temperature function using the chain rule.

$$n(\Omega_{m0},\Omega_{\Lambda 0},\sigma_8,w,z,kT)=\frac{\rho_{b0}}{3M\,kT}\frac{2.208m^p-1.534m^{2p}+2.385m^{3p}}{1+2.208m^p-0.767m^{2p}+0.795m^{3p}}\nu(kT)\,f(\nu(kT))\,. \quad (14)$$

Now

$$m=344.5\Omega_{m0}^2\exp[-0.088(1+1.192/\Omega_{m0})]\left[\frac{18\pi^2}{\Omega_{m0}\Delta(\Omega_{m0},\Omega_{\Lambda 0},w,z)}\right]^{1/2}\left(\frac{\beta_{TM}kT}{7.98\text{keV}(1+z)}\right)^{3/2} \quad (15)$$

Inserting the numerical constants yields our final function.

$$n(\Omega_{m0},\Omega_{\Lambda 0},\sigma_8,w,z,kT)=1.16x10^{-5}(h^{-1}\text{Mpc})^{-3}(\text{keV})^{-1}\beta_{TM}\Omega_{m0}\left(\frac{7.98\text{keV}}{\beta_{TM}kT}\right)^{5/2}$$

$$\frac{2.208m^p-1.534m^{2p}+2.385m^{3p}}{1+2.208m^p-0.767m^{2p}+0.795m^{3p}}\left[\frac{\Omega_{m0}\Delta(\Omega_{m0},\Omega_{\Lambda 0},w,z)}{18\pi^2}\right]^{1/2}(1+z)^{3/2}\nu(kT)\,f(\nu(kT)) \quad (16)$$

where we have used (15) to convert ν(M) to ν(kT). The coefficient of the equation in H00 analogous to (16) here should have been more general, i.e. $1.03x10^{-5}\ \beta_{TM}$. The value of $\beta_{TM}$ used in H00, 1.21, then yields the coefficient in H00.

In equation (16) we use the late – collapse approximation whereby the virialization epoch equals the epoch of observation. There has also been much discussion of the accuracy of this approximation. Mathiesen (2001) performed hydrodynamical simulations of 24 clusters, each imaged at 16 times in its history. He found no evidence for a relationship between present cluster temperature and formation epoch for those clusters that acquired 75 percent of their final mass since z = 0.6 because a cluster's temperature continually increases as it accretes mass. He concludes that the formation epoch is not an important variable in the analysis of cluster temperature functions. Enoki, Takahara & Fujita (2001) draw the same conclusion. They note that the virial temperature is $T_v \sim M_v/r_v \sim \rho_v^{1/3}M_v^{2/3}$, where all quantities have their usual meanings, and the evolution of the mass and density go in the opposite directions thus tending to cancel. Pierpaoli, Scott & White (2001) note that determining $\beta_{TM}$ from a sample of clusters, as we have done, either in simulations or in the real universe, already includes the effects of the formation redshift. Explicitly accounting for it as well would be double counting.

## 5. MEASURING COSMOLOGICAL PARAMETERS

Much of the previous work that used cluster evolution to constrain cosmology focused on two cosmological models. These were an open universe ($\Omega_{\Lambda 0}$ = 0) or a flat universe ($\Omega_{m0}$ + $\Omega_{\Lambda 0}$ = 1). Our formulation of these models had three free parameters, $\Omega_{m0}$ and two parameters specifying the shape and amplitude of the matter fluctuation power spectrum n and $\sigma_8$. In this paper we use the shape parameter Γ, which is a function of $\Omega_{m0}$ and other cosmological parameters. We adopt the standard model values for the other parameters as described in Section 4. The uncertainties in these other parameters propagate to smaller uncertainties in the parameters of interest than the statistical errors from the sample size or those from $\beta_{TM}$. Thus in our new formulation the models fitted previously had 2 parameters of interest.

We consider two models in this paper that are more general, both with three interesting parameters. The first has an arbitrary matter density and cosmological constant, with $\Omega_{\Lambda 0}$ the new parameter. It has become possible to constrain this model because Pierpaoli et al. (2001) recently derived the cluster average density Δ for it. The second model generalizes the dark energy in a flat universe by introducing an equation of state for this component with third parameter w: P = w ρ $c^2$ (Wang & Steinhardt 1998). If

w = -1, the dark energy is the cosmological constant, if $-1 < w < 0$ it is termed quintessence. Recall that an equation of state with w = 0 corresponds to cold dark matter and w = 1/3 is radiation. We give the details of these two models in the Appendix. The shape of the matter fluctuation power spectrum at cluster scales is the same for both of these models (Ma et al., 1999). To summarize, each model has 3 interesting parameters plus the normalization of the mass-temperature relation, which we marginalize over the range 0.7 – 1.0 assumed uniformly distributed.

We fit the temperature and redshift pairs in Tables 1 and 2 to these models using the maximum likelihood procedure of §3. The parameter space we explore is: $\Omega_{m0}$ = 0.2 (0.008) 0.8, $\Omega_{\Lambda 0}$ = 0.0 (0.02) 1.0, $\sigma_8$ = 0.4 (0.01) 0.9 and $\Omega_{m0}$ = 0.05 (0.01) 0.8, w = -1.0 (0.020) 0.0, $\sigma_8$ = 0.4 (0.01) 0.9 where the notation is minimum value (step size) maximum value. A total of 593,028 cases are evaluated for each of the general and the quintessence models for each value of $\beta_{TM}$. This range of parameters is a prior that we impose.

### 5.1 *Low Redshift Sample Comparison*

In Figure 3 we compare the constraints on $\sigma_8$ and $\Omega_{m0}$, the two best determined parameters, using the kT ≥ 3 keV low-redshift sample of Ikebe et al. (2002) analyzed with their method and using our low-redshift sample analyzed with our method but using the same value of $\beta_{TM}$ used by Ikebe et al. (1.07). The cutoff in our method for $\Omega_{m0} < 0.2$ is because Pierpaoli et al. (2001) did not derive $\Delta$ for that range. There is very good agreement. Figure 3 also shows a comparison of the constraints provided by the Ikebe et al. sample as determined by our method and their method, again with very good agreement. These comparisons demonstrate that the choice of low-redshift sample or the analysis method do not have a large impact on the final cosmological constraints. This result likely occurs because the low redshift temperature functions constructed from our sample and that of Ikebe et al. are nearly identical, as shown in Figure 6 of Ikebe et al. (2002).

### 5.2 *Results Not Greatly Dependent on $\beta_{TM}$*

There are some combinations of cosmological parameters whose constraints from our data do not depend greatly on $\beta_{TM}$. That is both the best fit values and error contours hardly change with the mass-temperature relation normalization. The uncertainties for these parameters do not depend on the details of the marginalization process over $\beta_{TM}$.

For the general cosmology model, the best fit parameters are $\Omega_{\Lambda 0}$ = 0.98, $\Omega_{m0}$ = 0.28 and $\sigma_8$ = 0.64. Figure 4 shows the 68% and 95% confidence constraints for two interesting parameters in the $\Omega_{\Lambda 0}$ - $\Omega_{m0}$ plane. An analysis that assumes that the cluster baryon fraction is the same for all clusters and does not evolve yields similar constraints (Allen, Schmidt & Fabian 2002; Ettori, Tozzi & Rosati, 2003; Vikhlinin et al. 2003a). We also show in Figure 4 the same confidence level constraints from the SNIa Hubble diagram (Tonry et al. 2003) and from the cosmic microwave background (Spergel et al. 2003). The cluster constraints are a band in this plane of different orientation to the latter two. All three bands intersect at approximately the same point in the plane; $\Omega_{m0}$ = 0.3, $\Omega_{\Lambda 0}$ = 0.7 is consistent with cluster and microwave background at the ~1$\sigma$ level and supernovae data at the ~2$\sigma$ level.

For the quintessence model, the best-fit parameters are w = -0.42, $\Omega_{m0}$ = 0.33 and $\sigma_8$ = 0.55. Figure 5 shows the 68% and 95% confidence constraints for two interesting parameters in the w - $\Omega_{m0}$ plane. The cluster constraints are very similar to the large scale structure results in Figure 1 of Perlmutter, Turner & White (1999). Considering only a single interesting parameter, w is different from –1 at 98% confidence. Figure 6 shows the 1$\sigma$ constraints provided by our clusters, REFLEX (low redshift) clusters (Schuecker et al., 2003b), supernovae (Turner & Reiss, 2002), CMB (Spergel et al., 2003), and gravitational lensing of radio sources (Chae et al., 2002). The evolution of the cluster gas mass, assuming the baryon fraction does not change, yields similar constraints (Vikhlinin et al., 2003b, c, d). All results are consistent with w = -0.8 and $\Omega_{m0}$ = 0.3.

### 5.3 *Results Dependent on $\beta_{TM}$*

Cluster number density evolution can provide tight statistical constraints on the normalization of the power spectrum $\sigma_8$. Conversely, systematic effects become important as well. This is an unfortunate situation, since the final error bars on $\sigma_8$ depend on the details of the marginalization over uninteresting parameters.

We show in Figure 7 the 68% and 95% confidence constraints on $\sigma_8$ vs. $\Omega_{m0}$ for 3 interesting parameters for the average $\beta_{TM}$ from empirical determinations (0.72) and from hydrodynamical simulations (1.03). We also show the 68% and 95% confidence standard model values. This figure shows clearly that the statistical errors on $\sigma_8$ are much smaller than the systematic uncertainties on $\beta_{TM}$. The figure also shows that this is not yet the case for $\Omega_{m0}$.

In Figure 8 we compare the cluster constraints on $\sigma_8$ and. $\Omega_{m0}$ with those from other data sets. We show the 95% confidence constraints from the CMB (Bridle et al. 2003; the 68% confidence contour was not given). The cluster constraints agree with those from the CMB at ~1σ, showing that the fluctuation in the CMB really do grow to form the clusters seen today. We also show in Figure 8 the 68% and 95% confidence constraints from the analysis of cosmic shear from weak lensing (Figure 3a from Hoekstra, Yee & Gladders, 2002). We use this lensing analysis because it is the mean of seven such analyses summarized by Pierpaoli et al. (2003) and it reports the results obtained with very weak priors. The values $\Omega_{m0}$ = 0.30, $\sigma_8$ = 0.8 are consistent with all three at the ~1σ level. The $\sigma_8$ - $\Omega_{m0}$ constraints from clusters for the quintessence model are virtually the same as those shown in Figure 8.

Several authors (Voit, 2000; Wu, 2001; Pierpaoli et al., 2001, 2003; Viana et al., 2003) have noted the strong dependence of cosmological constraints from cluster evolution on the mass – temperature relation normalization. We have collected in Table 4 the most recent determinations of $\sigma_8$ using clusters and plotted them against $\beta_{TM}$ in Figure 9. There is indeed a correlation between these two parameters that is significant at the 99.2% confidence level. The best fitting linear regression to the unweighted data is $\sigma_8(0.3) = 0.42 + 0.38\beta_{TM}$. Much of the scatter in the reported values of $\sigma_8$ is due simply to differences in $\beta_{TM}$. Viana et al. (2003) give a relation between $\sigma_8$ and the normalization of the mass - temperature relation, their equation 9. Converting this relation to our notation yields $\sigma_8(0.3) = 0.40 + 0.48\,\beta_{TM}^{1.25}$, which provides a reasonable description of the data in Figure 9 although perhaps somewhat high.

This uncertainty in the value of the mass – temperature relation normalization is the most pressing issue thwarting precision cosmology using cluster luminosity or temperature evolution, at least for measurements of $\sigma_8$. The uncertainty translates to a $\sigma_8$ uncertainty of ±10%. If we were to adopt the standard model values for $\sigma_8$, then we could determine $\beta_{TM}$ using the linear regression relation given above. Of course cluster data would then not test the standard model. Using this procedure, we find the standard model value of $\sigma_8(0.3)$ = 0.79 implies $\beta_{TM}$ = 0.97. This value is close to those determined through simulations.

This paper has benefited from discussions with H. Böhringer, S. Bridle and Y. Ikebe. I thank Y. Suto for convincing me to use the CDM power spectrum and D. Koo for sharing his redshifts in the MS0015.9 field. I gratefully acknowledge financial support from NASA grant NAG5-9166.

TABLE 1
LOW-REDSHIFT SAMPLE

| Name | z | F(2,10) ($10^{-13}$ ergs $cm^{-2}$ $s^{-1}$) | kT (keV) | Vsea ($h^{-3}$ $Mpc^3$) | n(>kT) ($h^3$ $Mpc^{-3}$) |
|---|---|---|---|---|---|
| A754 | 0.0543 | 8.5 | 9.83±0.27 | 1.65E8 | 0.00E0 |
| A2142 | 0.0894 | 7.5 | 9.02±0.32 | 1.17E8 | 6.08E-9 |
| A401 | 0.0736 | 5.9 | 8.68±0.17 | 1.00E8 | 1.47E-8 |
| A1656 | 0.0232 | 32.0 | 8.67±0.17 | 1.10E7 | 2.37E-8 |
| A3266 | 0.0590 | 5.9 | 8.34±0.17 | 8.63E7 | 3.28E-8 |
| A2029 | 0.0766 | 7.5 | 8.22±0.21 | 8.35E7 | 4.44E-8 |
| A3571 | 0.0397 | 11.5 | 7.24±0.09 | 5.13E7 | 5.64E-8 |
| A3667 | 0.0556 | 6.7 | 7.13±0.14 | 5.07E7 | 7.59E-8 |
| A2256 | 0.0581 | 5.2 | 6.96±0.11 | 4.38E7 | 9.56E-6 |
| A399 | 0.0723 | 3.4 | 6.80±0.17 | 3.88E7 | 1.18E-7 |
| A478 | 0.0882 | 6.6 | 6.58±0.26 | 3.53E8 | 1.44E-7 |
| A1651 | 0.0846 | 3.7 | 6.21±0.18 | 2.47E7 | 1.73E-7 |
| A85 | 0.0556 | 6.4 | 5.92±0.11 | 2.32E7 | 2.13E-7 |
| A1795 | 0.0630 | 5.3 | 5.80±0.07 | 1.92E7 | 2.56E-7 |
| A119 | 0.0442 | 3.0 | 5.62±0.12 | 1.62E7 | 3.08E-7 |
| A3558 | 0.0479 | 4.2 | 5.53±0.09 | 1.62E7 | 3.70E-7 |
| A3562 | 0.0502 | 3.5 | 5.16±0.16 | 1.14E7 | 4.32E-7 |
| A2147 | 0.0353 | 3.3 | 4.50±0.07 | 6.26E7 | 5.19E-7 |
| A2199 | 0.0299 | 7.1 | 4.27±0.04 | 4.75E6 | 6.79E-7 |
| A496 | 0.0328 | 5.7 | 4.08±0.04 | 3.98E6 | 8.90E-7 |
| A1367 | 0.0213 | 3.5 | 3.64±0.10 | 2.48E6 | 1.14E-6 |
| A3526 | 0.0103 | 11.2 | 3.42±0.02 | 1.68E6 | 1.54E-6 |
| A1060 | 0.0124 | 4.4 | 3.19±0.03 | 1.08E6 | 2.14E-6 |
| C0336 | 0.0349 | 4.7 | 3.08±0.03 | 1.04E6 | 3.07E-6 |
| Virgo | 0.0037 | 30.0 | 2.40±0.01 | | |

Notes—Temperature errors are 68% confidence level. Search volumes and temperature functions assume $\Omega_{m0} = 0.3$, $\Omega_{\Lambda 0} = 0.7$.

TABLE 2
HIGH-REDSHIFT SAMPLE

| | | $F_{det}(0.3,3.5)$ ($10^{-13}$ ergs cm$^{-2}$ s$^{-1}$) | | $L_{44}(2,10)$ | $L_{44}(0.5,2.0)$ | kT | Vsea | n(>kT) |
|---|---|---|---|---|---|---|---|---|
| MS | z | EMSS | ASCA | ASCA | ASCA | (keV) | ($h^{-3}$ Mpc$^3$) | ($h^3$ Mpc$^{-3}$) |
| 0451.6 | 0.5392 | 9.5±1.7 | 12.4±0.2 | 30.6±0.6 | 13.3±0.3 | $10.3^{+0.9}_{-0.8}$ | 4.53E8 | 0.00E-0 |
| 1054.4 | 0.8309 | 2.1±0.3 | 3.7±0.5 | 20.6±2.8 | 9.0±1.2 | $10.1^{+1.8}_{-1.5}$ | 4.42E8 | 2.21E-9 |
| 1137.5 | 0.7843 | 1.9±0.4 | 1.9±0.4 | 9.5±0.8 | 4.1±0.3 | $9.5^{+2.8}_{-2.0}$ | 4.08E8 | 4.47E-9 |
| 0015.9 | 0.5466 | 7.1±0.8 | 14.1±0.3 | 34.1±0.6 | 15.9±0.3 | $8.9^{+0.6}_{-0.6}$ | 3.71E8 | 6.92E-9 |
| 1008.1 | 0.3062 | 5.9±0.8 | 9.2±0.3 | 8.7±0.3 | 4.0±0.1 | $8.2^{+1.2}_{-1.1}$ | 3.19E8 | 9.61E-9 |
| 2053.7 | 0.583 | 2.5±0.6 | 2.4±0.2 | 6.6±0.5 | 3.1±0.2 | $8.1^{+3.7}_{-2.2}$ | 3.13E8 | 1.28E-8 |
| 0418.3 | 0.350 | 1.5±0.3 | 2.7±0.1 | 2.7±0.1 | 1.4±0.1 | $7.0^{+1.4}_{-1.4}$ | 2.21E8 | 1.59E-8 |
| 1358.4 | 0.3290 | 12.2±2.1 | 11.0±0.3 | 10.2±0.3 | 5.4±0.2 | $6.9^{+0.5}_{-0.5}$ | 2.12E8 | 2.05E-8 |
| 1621.5 | 0.4274 | 3.4±0.6 | 5.2±0.2 | 7.6±0.3 | 4.0±0.2 | $6.6^{+0.9}_{-0.8}$ | 1.84E8 | 2.52E-8 |
| 0353.6 | 0.320 | 6.3±1.3 | 7.8±0.3 | 6.8±0.3 | 3.7±0.1 | $6.5^{+1.0}_{-0.8}$ | 1.73E8 | 3.06E-8 |
| 1426.4 | 0.320 | 4.4±0.6 | 5.3±0.2 | 4.6±0.2 | 2.5±0.1 | $6.4^{+1.0}_{-1.2}$ | 1.66E8 | 3.64E-8 |
| 1241.5 | 0.549 | 4.2±1.1 | 4.7±0.3 | 10.7±0.6 | 5.8±0.3 | $6.1^{+1.4}_{-1.1}$ | 1.42E8 | 4.24E-8 |
| 1147.3 | 0.303 | 3.0±0.7 | 5.3±0.2 | 4.1±0.2 | 2.3±0.1 | $6.0^{+1.0}_{-0.7}$ | 1.32E8 | 4.94E-8 |
| 0821.5 | 0.347 | 1.4±0.4 | 1.7±0.2 | 1.5±0.1 | 0.93±0.08 | $5.9^{+1.5}_{-1.5}$ | 1.27E8 | 5.70E-8 |
| 0811.6 | 0.312 | 2.6±0.5 | 3.5±0.2 | 2.6±0.2 | 1.7±0.1 | $4.9^{+1.0}_{-0.6}$ | 5.71E7 | 6.49E-8 |
| 2137.3 | 0.313 | 19.3±2.6 | 20.9±0.4 | 15.3±0.3 | 10.0±0.2 | $4.9^{+0.3}_{-0.3}$ | 5.61E7 | 8.24E-8 |
| 1208.7 | 0.340 | 2.2±0.3 | | | | $4.4^{+0.7}_{-0.7}$ [a] | 3.38E7 | 1.00E-7 |
| 1532.5 | 0.320 | 2.0±0.5 | 2.5±0.2 | 1.7±0.1 | 1.3±0.1 | $4.1^{+0.7}_{-0.6}$ | 2.35E7 | 1.30E-7 |
| 1224.7 | 0.3255 | 5.3±1.1 | 4.2±0.2 | 3.0±0.2 | 2.2±0.1 | $4.1^{+0.7}_{-0.5}$ | 2.25E7 | 1.72E-7 |
| | | | | | | | | |
| 0302.7 | 0.4246 | 3.8±0.4 | 4.2±0.3 | 4.8±0.3 | 3.5±0.2 | $4.4^{+0.8}_{-0.6}$ | | |
| 0302.5 | 0.4249 | 2.2±0.4 | 4.1±0.2 | 6.1±0.3 | 3.1±0.2 | $7.9^{+2.0}_{-1.5}$ | | |
| 1512.4 | 0.3726 | 4.5±1.0 | 4.5±0.3 | 3.8±0.3 | 3.2±0.2 | $3.4^{+0.4}_{-0.4}$ | | |
| 0147.8 | 0.373 | 1.5±0.3 | 0.55±0.24 | | | [b] | | |

Notes—All errors are at the 68% confidence level. Luminosities assume $\Omega_{m0} = 1$ and $H_0 = 50$ km s$^{-1}$ Mpc$^{-1}$. Search volumes and temperature functions assume $\Omega_{m0} = 0.3$, $\Omega_{\Lambda 0} = 0.7$.

[a] Temperature from L – T relation with high redshift sample average fractional temperature error.

[b] Insufficient counts to fit a temperature.

TABLE 3

DETERMINATION OF THE MASS-TEMPERATURE PARAMETER $\beta_{TM}$

| Method | $\beta_{TM}$ | $\Omega_{m0}$ | $\Omega_b$ | $\Omega_{\Lambda 0}$ | h | $\sigma_8$ | Objects | Reference |
|---|---|---|---|---|---|---|---|---|
| Simulation | 1.17 | 1.00 | 0.10 | 0.00 | 0.50 | 0.59, 0.63 | 42 | 1 |
| Simulation | 0.88 | 0.20 | 0.10 | 0, 0.8 | 0.50 | 1.00 | 16 | 1 |
| Simulation | 1.15±0.05 | 1.00 | 0.05 | 0.00 | 0.50 | 0.50 | 20 | 2 |
| Simulation | 1.07±0.04 | 0.30 | 0.04 | 0.70 | 0.70 | 1.05 | 30 | 3 |
| Simulation | 1.29 | 1.00 | 0.06 | 0.00 | 0.50 | 1.05 | 25 | 4 |
| Simulation | 1.36 | 1.00 | 0.10 | 0.00 | 0.65 | 0.67 | 25 | 4 |
| Simulation | 1.36 | 1.00 | 0.075 | 0.00 | 0.50 | 0.70 | 25 | 4 |
| Simulation | 1.31 | 0.40 | 0.06 | 0.00 | 0.65 | 0.75 | 25 | 4 |
| Simulation | 0.96 | 0.30 | 0.03 | 0.70 | 0.70 | 1.00 | 615 | 5 |
| Simulation | 1.02 | 1.00 | 0.06 | 0.00 | | 0.60 | ~400 | 6 |
| Simulation | 1.03 | 0.35 | 0.04 | 0.65 | 0.70 | 0.90 | 427 | 7 |
| Simulation | 0.94±0.15 | 0.35 | 0.038 | 0.65 | 0.71 | 0.7, 0.9 | 43 | 8 |
| Gravitational lens | 1.02±0.11 | 1.00 | | 0.00 | 0.50 | | 8 | 9 |
| Spatially resolved temperature | $0.76^{+0.28}_{-0.20}$ | 1.00 | | | 1.00 | | 11 | 10 |
| Spatially resolved temperature | $0.76^{+0.11}_{-0.05}$ | 1.00 | | 0.00 | 0.50 | | 24 | 11 |
| Gravitational lens | $0.85^{+0.19}_{-0.26}$ | 0.30 | | 0.70 | 1.00 | | 15 | 12 |
| Spatially resolved temps Gravitational lens | 0.70±0.04 | 0.30 | | 0.70 | 0.50 | | 13 | 13 |

REFERENCES – (1) Evrard, Metzler & Navarro 1996; (2) Pen 1998; (3) Eke, Navarro & Frenk 1998; (4) Bryan & Norman 1998; (5) Yoshikawa, Jing & Suto 2000; (6) Thomas et al. 2001; (7) Muanwong et al. 2001; (8) Viana, P. T. P. et al. 2003; (9) Hjorth, Oukbir & van Kampen 1998; (10) Horner, Mushotzky & Scharf 1999; (11) Finoguenov, Reiprich & Böhringer 2001 (kT > 3 keV); (12) Sheldon et al. (2001); (13) Allen et al. (2002b)

TABLE 4

RECENT CLUSTER DETERMINATIONS OF $\sigma_8$ WITH $\Omega_{m0}$ = 0.3

| Reference | $\beta_{TM}$ | $\sigma_8(\Omega_{m0} = 0.3)$ | Method [a] | Sample |
|---|---|---|---|---|
| Viana et al. (2002) | $0.58 \pm 0.09$ | $0.61 \pm 0.05$ | LF Local | REFLEX |
| This paper | 0.72 | $0.62 \pm 0.04$ | TF Evolution | HA + EMSS |
| Reiprich & Boehringer (2002) | - | $0.68 \pm 0.04$ | MF Local | HIFLUGS |
| Allen et al. (2003a) | $0.70 \pm 0.04$ | $0.69 \pm 0.04$ | LF Local | REFLEX + BCS |
| Bahcall et al. (2003) | - | $0.72 \pm 0.06$ | MF Local | SDSS |
| Rosati et al. (2002) | 1.15 | $0.72 \pm 0.02$ | LF Evolution | RDCS |
| Schuecker et al. (2003a) | - | $0.76 \pm 0.01$ | LF+PS Local | REFLEX |
| Seljak (2002) | $0.77 \pm 0.10$ | $0.76 \pm 0.06$ | TF Local | Pierpaoli et al. |
| This paper | 1.03 | $0.76 \pm 0.04$ | TF Evolution | HA + EMSS |
| Ikebe et al. (2002) | 1.07 | $0.77 \pm 0.03$[b] | TF Local | Ikebe et al., kT>3 |
| Pierpaoli et al. (2003) | $0.81^{+0.14}_{-0.06}$ | $0.77 \pm 0.05$ | TF Local | HIFLUGCS, kT>3 |
| Pierpaoli et al. (2003) | $0.81^{+0.14}_{-0.06}$ | $0.78 \pm 0.06$ | LF Local | REFLEX |
| Viana et al. (2003) | $0.94 \pm 0.15$ | $0.84^{+0.16}_{-0.03}$ | TF Local | Ikebe et al., kT>2 |
| Wu (2001) | $1.28 \pm 0.23$ | $0.91 \pm 0.11$ | TF Local | HA, kT>6.2 |
| Pierpaoli et al. (2001) | 1.09 | $1.02 \pm 0.07$ | TF Local | Pierpaoli et al. |

Notes: [a] LF = luminosity function, TF = temperature function MF = mass function PS = power spectrum
[b] Converted from Press-Schechter mass function using a factor 0.88 derived from a fit to our the data.

APPENDIX

We continue the appendix from H00 to the more general cosmological models considered in this paper. The quantities given here are needed for our theoretical analysis and for the maximum likelihood fits to our temperature data. Recalling, $\delta_c(\Omega_{m0}, \Omega_{\Lambda 0}, z)$ is the critical overdensity for collapse at redshift z; $\Delta(\Omega_{m0}, \Omega_{\Lambda 0}, z)$ is the ratio of the average cluster mass density to that of the background density for a cluster which collapses at redshift z; $D(\Omega_{m0}, \Omega_{\Lambda 0}, z)$ is the growth factor from redshift z; $D_L(\Omega_{m0}, \Omega_{\Lambda 0}, z)$ is the luminosity distance to redshift z and the angular diameter distance is $D_L(\Omega_{m0}, \Omega_{\Lambda 0}, z)/(1+z)^2$; V is the volume and $d\Omega$ is the solid angle surveyed. Quantities with subscript zero are evaluated at the present epoch.

We note one error we have found in the appendix of H00. The equation between (A10) and (A11) for the growth factor, which somehow escaped being numbered, should not have a 2 in the denominator if it is to go to 1/(1+z) in the limit that $\Omega_{m0} \to 1$. This error has no effect on any result in H00 since the growth factor always enters as a ratio of factors at two redshifts.

A4. GENERAL MODEL

$$x_\Lambda \equiv \left(\frac{\Omega_{\Lambda 0}}{\Omega_{m0}}\right)^{1/3} \frac{1}{1+z} \tag{A24}$$

$$x_k \equiv \frac{1-\Omega_{m0}-\Omega_{\Lambda 0}}{\Omega_{m0}} \frac{1}{1+z} \equiv \frac{\Omega_{k0}}{\Omega_{m0}} \frac{1}{1+z} \tag{A25}$$

$$\Omega_m(z) = \frac{1}{1+x_k+x_\Lambda^3} \tag{A26}$$

$$\Omega_\Lambda(z) = \frac{x_\Lambda^3}{1+x_k+x_\Lambda^3} \tag{A27}$$

$$\Omega_k(z) = \frac{x_k}{1+x_k+x_\Lambda^3} \tag{A28}$$

From these it follows that $x_\Lambda = [\Omega_\Lambda(z)/\Omega_m(z)]^{1/3}$ and $x_k = (1 - \Omega_m(z) - \Omega_\Lambda(z))/ \Omega_m(z)$.

$$\delta_c(\Omega_{m0}, \Omega_{\Lambda 0}, z) \approx 1.686 \tag{A29}$$

We expect a few percent variations in this parameter due to the cosmology, but this amount is similar to the errors on some of the fits below, so we neglect it.

$$\Delta(\Omega_{m0}, \Omega_{\Lambda 0}, z) = \sum_{i,j=0}^{4} c_{ij} (\Omega_m(z) - 0.2)^i \Omega_\Lambda^j(z) \tag{A30}$$

valid for $0.2 \le \Omega_{m0}(z) \le 1.1$ and $0 \le \Omega_{\Lambda 0}(z) \le 1$ (Pierpaoli, Scott, & White 2001, eq. [10]). Here the $c_{ij}$ are

Table A1

$c_{ij}$

| i | j = 0 | 1 | 2 | 3 | 4 |
|---|---|---|---|---|---|
| 0 | 546.67 | -137.82 | 94.083 | -204.68 | 111.51 |
| 1 | -1745.6 | 627.22 | -1175.2 | 2445.7 | -1341.7 |
| 2 | 3928.8 | -1519.3 | 4015.8 | -8415.3 | 4642.1 |
| 3 | -4384.8 | 1748.7 | -5362.1 | 11257. | -6218.2 |
| 4 | 1842.3 | -765.53 | 2507.7 | -5210.7 | 2867.5 |

$$D(\Omega_{m0},\Omega_{\Lambda 0},z)=\frac{x_\Lambda}{x_{\Lambda 0}}\sqrt{1+x_k+x_\Lambda^3}\int_0^1 dy[1+x_k y^{2/5}+x_\Lambda^3 y^{6/5}]^{-3/2} \qquad \text{(A31)}$$

(Carroll, Press, & Turner 1992, eq. [28]; Pierpaoli et al. 2001, eq. [3]).

$$D_L=\frac{c(1+z)}{H_0\sqrt{\Omega_{k0}}}\sinh\int_0^z dy\sqrt{\Omega_{k0}}\left[\Omega_{m0}(1+y)^3+\Omega_{k0}(1+y)^2+\Omega_{\Lambda 0}\right]^{-1/2} \qquad \Omega_{k0}>0 \quad \text{(A32a)}$$

$$D_L=\frac{c(1+z)}{H_0}\int_0^z dy\left[\Omega_{m0}(1+y)^3+\Omega_{\Lambda 0}\right]^{-1/2} \qquad \Omega_{k0}=0 \quad \text{(A32b)}$$

$$D_L=\frac{c(1+z)}{H_0\sqrt{-\Omega_{k0}}}\sin\int_0^z dy\sqrt{-\Omega_{k0}}\left[\Omega_{m0}(1+y)^3+\Omega_{k0}(1+y)^2+\Omega_{\Lambda 0}\right]^{-1/2} \qquad \Omega_{k0}<0 \quad \text{(A32c)}$$

(Carroll et al. 1992, eqs. [23] and [25]; Yoshii 1995, eq. [12]).

$$\frac{dV(\Omega_{m0},\Omega_{\Lambda 0},\le z)}{d\Omega}=\left(\frac{c}{H_0}\right)^3\frac{1}{2\Omega_{k0}}\left[\frac{H_0 D_L}{c(1+z)}\sqrt{1+\Omega_{k0}\left(\frac{H_0 D_L}{c(1+z)}\right)^2}-\frac{1}{\sqrt{\Omega_{k0}}}\sinh^{-1}\left(\frac{H_0 D_L\sqrt{\Omega_{k0}}}{c(1+z)}\right)\right]$$

$$\Omega_{k0}>0 \quad \text{(A33a)}$$

$$\frac{dV(\Omega_{m0},\Omega_{\Lambda 0},\le z)}{d\Omega}=\frac{1}{3}\left(\frac{D_L}{1+z}\right)^3 \qquad \Omega_{k0}=0 \quad \text{(A33b)}$$

$$\frac{dV(\Omega_{m0},\Omega_{\Lambda 0},\le z)}{d\Omega}=\left(\frac{c}{H_0}\right)^3\frac{1}{2\Omega_{k0}}\left[\frac{H_0D_L}{c(1+z)}\sqrt{1+\Omega_{k0}\left(\frac{H_0D_L}{c(1+z)}\right)^2}-\frac{1}{\sqrt{-\Omega_{k0}}}\sin^{-1}\left(\frac{H_0D_L\sqrt{-\Omega_{k0}}}{c(1+z)}\right)\right]$$

$\Omega_{k0} < 0$ (A33c)

(Carroll et al. 1992, eqs. [23] and [27]).

$$\frac{d^2V(\Omega_{mo},\Omega_{\Lambda 0},z)}{d\Omega dz}=\frac{c}{H_0}\frac{D_L^2(\Omega_{m0},\Omega_{\Lambda 0},z)}{(1+z)^2\sqrt{\Omega_{m0}(1+z)^3+\Omega_{k0}(1+z)^2+\Omega_{\Lambda 0}}} \quad \text{(A34)}$$

(Yoshii 1995, eq. [14]).

A5. QUINTESSENCE MODEL ($\Omega_{m0} + \Omega_{Q0} = 1$, $\Omega_{\Lambda 0} = 0$)

$$x\equiv\frac{(\Omega_{m0}^{-1}-1)^{-1/3w}}{1+z} \quad \text{(A35)}$$

$$\Omega_m(z)=\frac{1}{1+x^{-3w}} \quad \text{(A36)}$$

$$\delta_c(\Omega_{m0},w,z)=1.686\Omega_{m0}^{0.037/(1-w)^{2.7}} \quad \text{(A37)}$$

(Schuecker et al., 2003b).

$$\Delta(\Omega_{m0},w,z)=\zeta\left(\frac{R_{ta}}{R_{vir}}\right)^3\left(\frac{1+z_{ta}}{1+z}\right)^3 \quad \text{(A38)}$$

$$\zeta=\left(\frac{3\pi}{4}\right)^2\left[\Omega_m(z_{ta})\right]^{-0.79+0.26\Omega_m(z_{ta})-0.06w} \quad \text{(A39)}$$

$$\frac{R_{ta}}{R_{vir}}=\frac{2+2\zeta^{-1}(\Omega_m^{-1}(z_{ta})-1)-3\zeta^{-1}(\Omega_m^{-1}(z)-1)((1+z)/(1+z_{ta}))^3}{1-\zeta^{-1}(\Omega_m^{-1}(z)-1)((1+z)/(1+z_{ta}))^3} \quad \text{(A40)}$$

(Wang & Steinhardt 1998, egs. [5], [A11], [7]). Recall that the cluster collapsed at redshift z, $z_{ta}$ is defined implicitly by

$$\int_0^{1/(1+z)}dyy^{1/2}\left[\Omega_{m0}+(1-\Omega_{m0})y^{-3w}\right]^{-1/2}=2\int_0^{1/(1+z_{ta})}dyy^{1/2}\left[\Omega_{m0}+(1-\Omega_{m0})y^{-3w}\right]^{-1/2} \quad \text{(A41)}$$

$$D(\Omega_{m0},w,z)\approx\frac{1}{1+z}\exp\left[\int_0^z dy(1-\Omega_m^{\alpha(y)}(y))/(1+y)\right] \quad \text{(A42)}$$

with $$\alpha(y) \approx \frac{3}{5 - w/(1-w)} + \frac{3}{125}\frac{(1-w)(1-3w/2)}{(1-6w/5)^3}(1-\Omega_m(y)) \tag{A43}$$

(Wang & Steinhardt 1998, eqs. [13] and [14]).

$$D_L(\Omega_{m0}, w, z) = \frac{c}{H_0}(1+z)\int_0^z dy(\Omega_{m0}(1+y)^3 + (1-\Omega_{m0})(1+y)^{3(1+w)})^{-1/2} \tag{A44}$$

$$\frac{d^2V(\Omega_{m0}, w, z)}{d\Omega dz} = \frac{c}{H_0}\frac{D_L^2(\Omega_{m0}, w, z)}{(1+z)^2\sqrt{\Omega_{m0}(1+z)^3 + (1-\Omega_{m0})(1+z)^{3(1+w)}}} \tag{A45}$$

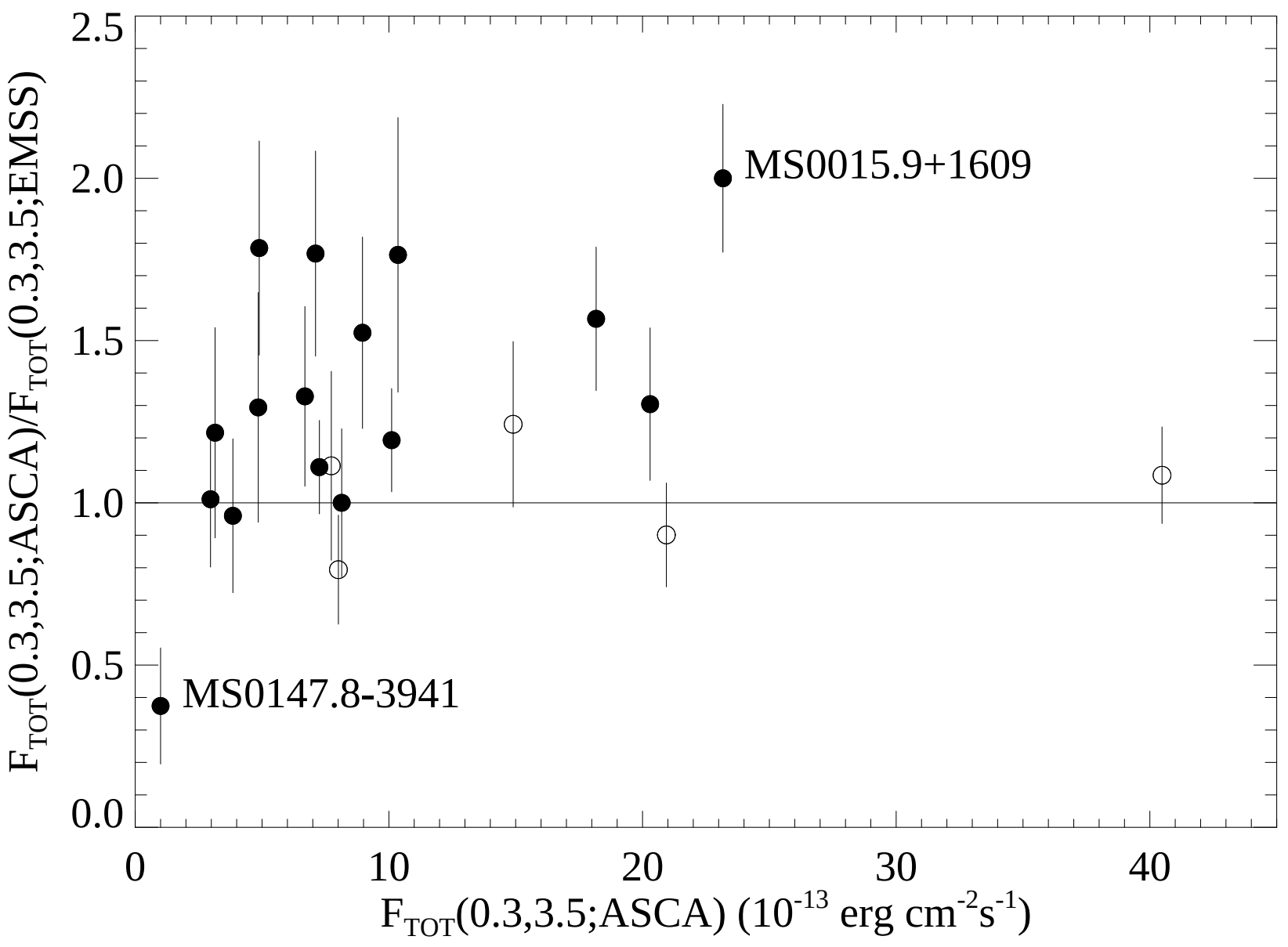

Figure 1. Ratio of ASCA to EMSS total fluxes plotted as a function of the ASCA total flux. The open circles are for clusters with no obvious point sources in their HRI images within 6.125′ of the cluster center.

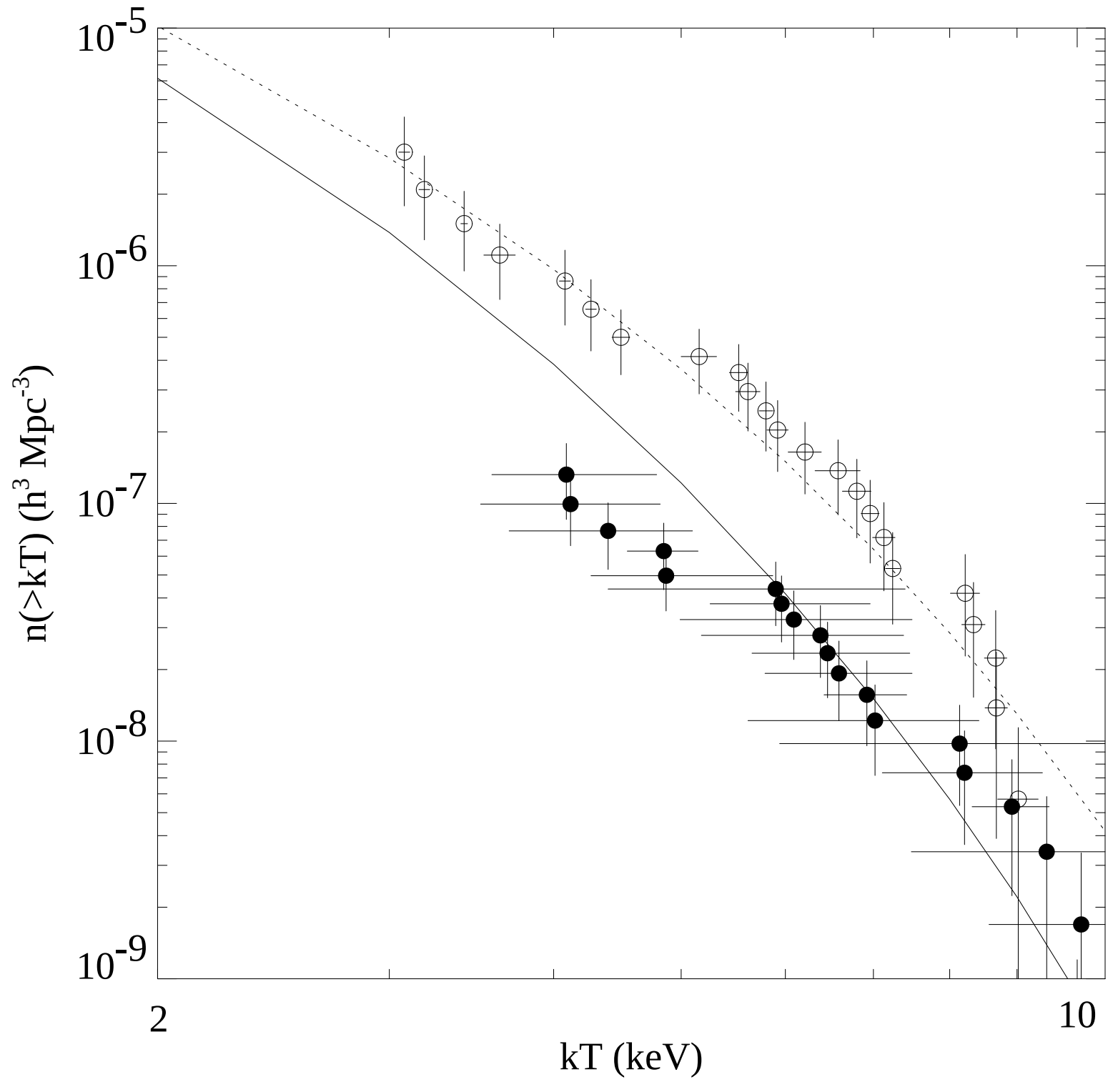

Figure 2. The open (filled) circles and the dotted (solid) lines are the low (high) redshift integral cluster temperature function observations and best fits respectively. Both assume $\Omega_{m0} = 0.28$ and $\Omega_{\Lambda 0} = 0.98$, the best fitting values.

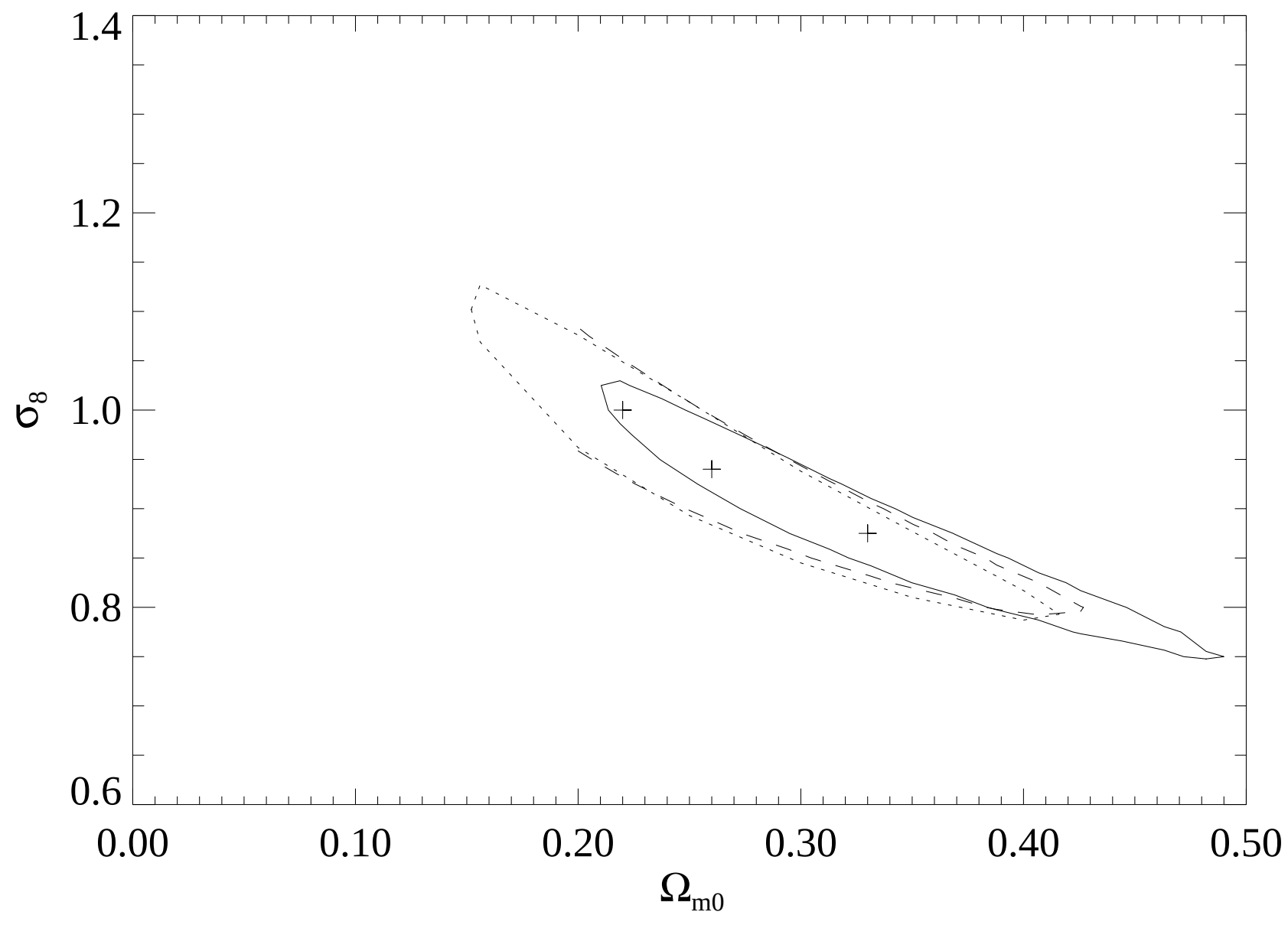

Figure 3. The 90% confidence (1.6 $\sigma$) constraints on $\sigma_8$ and $\Omega_{m0}$ for two interesting parameters (i.e. $\Delta S = 4.61$) from low-redshift kT > 3 keV samples only, assuming $\Omega_{\Lambda 0} + \Omega_{m0} = 1$. Crosses are the best fit points. The dotted line is the analysis by Ikebe et al. of their sample. The solid line is our analysis using our code of the Ikebe et al. sample; the dashed line is our analysis of our sample. This figure shows that the particular analysis of the particular low-redshift sample has little effect on the cosmological constraints.

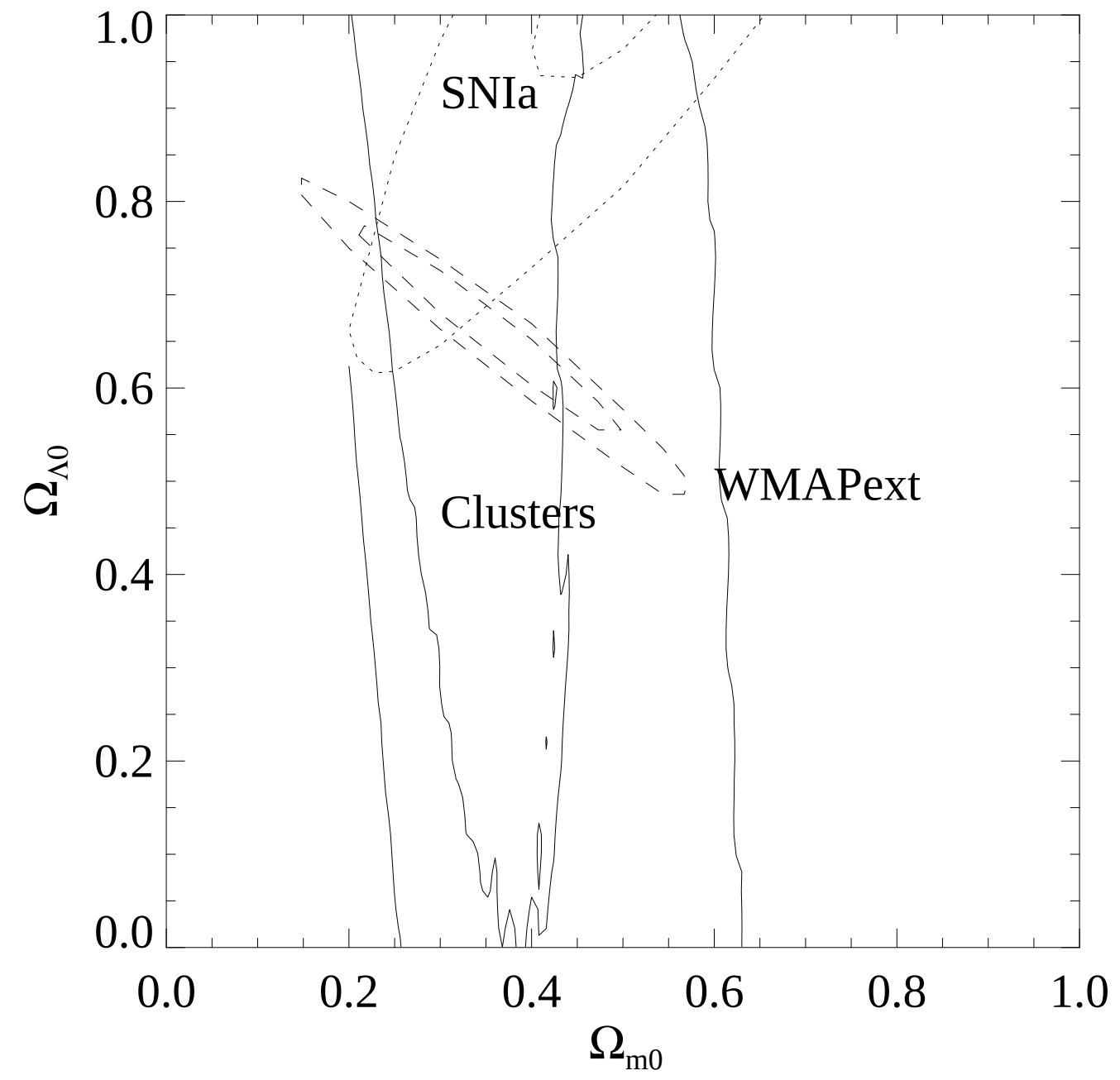

Figure 4. The 1 and 2 $\sigma$ constraints on $\Omega_{\Lambda 0}$ and $\Omega_{m0}$ from cluster temperature evolution (solid), cosmic microwave background assuming $h > 0.5$ (dashed; Spergel et al. (2003) and supernovae (dotted; Tonry et al. (2003). The cluster constraints are for 2 interesting parameters (i.e. $\Delta S = 2.3$, 6.2) and have been marginalized over $\beta_{TM}$.

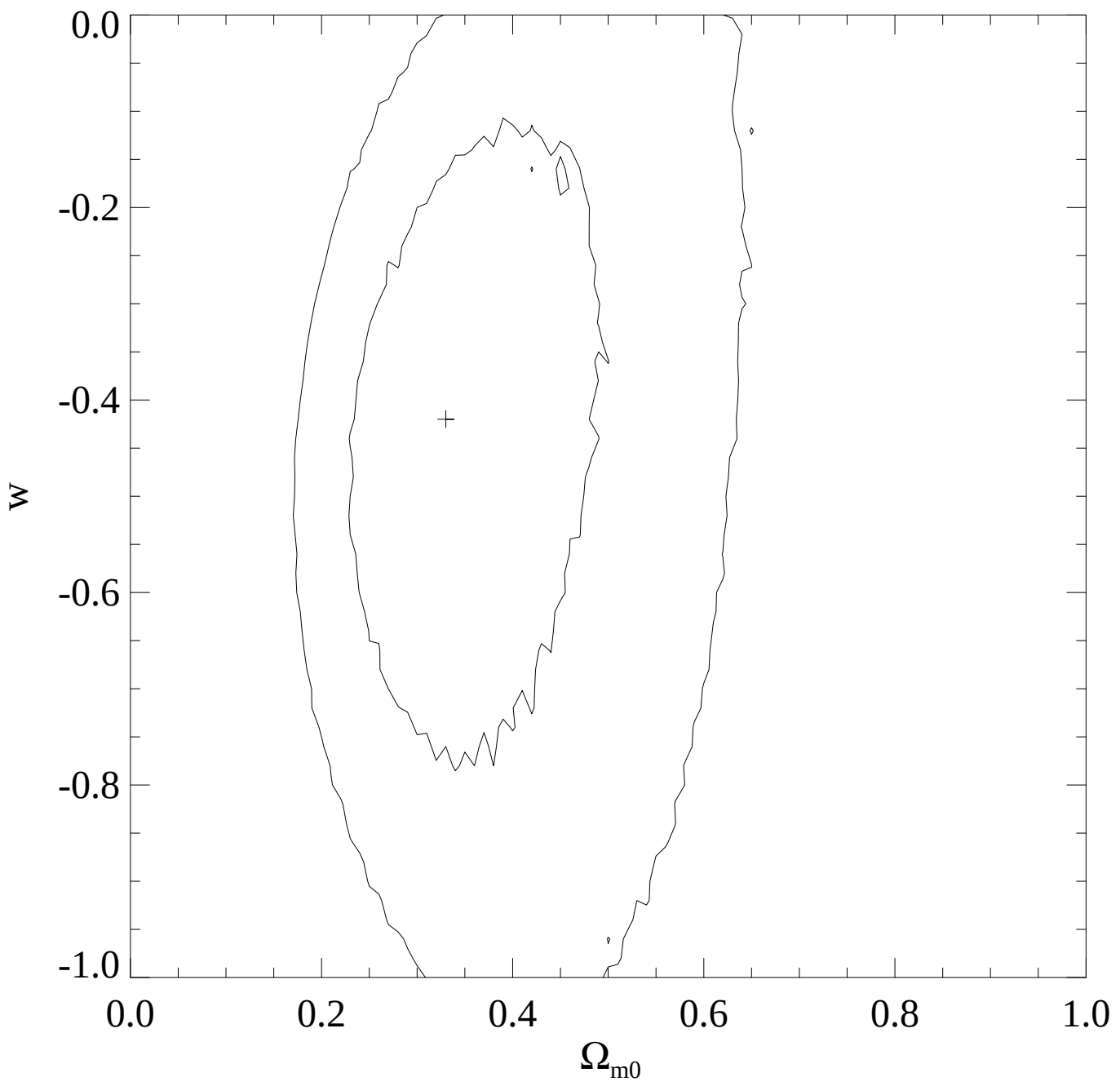

Figure 5. Constraints on w and $\Omega_{m0}$ are shown as 1 and 2 $\sigma$ contours from cluster temperature evolution for two interesting parameters ((i.e. $\Delta S = 2.3, 6.2$). The constraints have been marginalized over $\beta_{TM}$ from 0.7 to 1.0, assumed uniformly distributed.

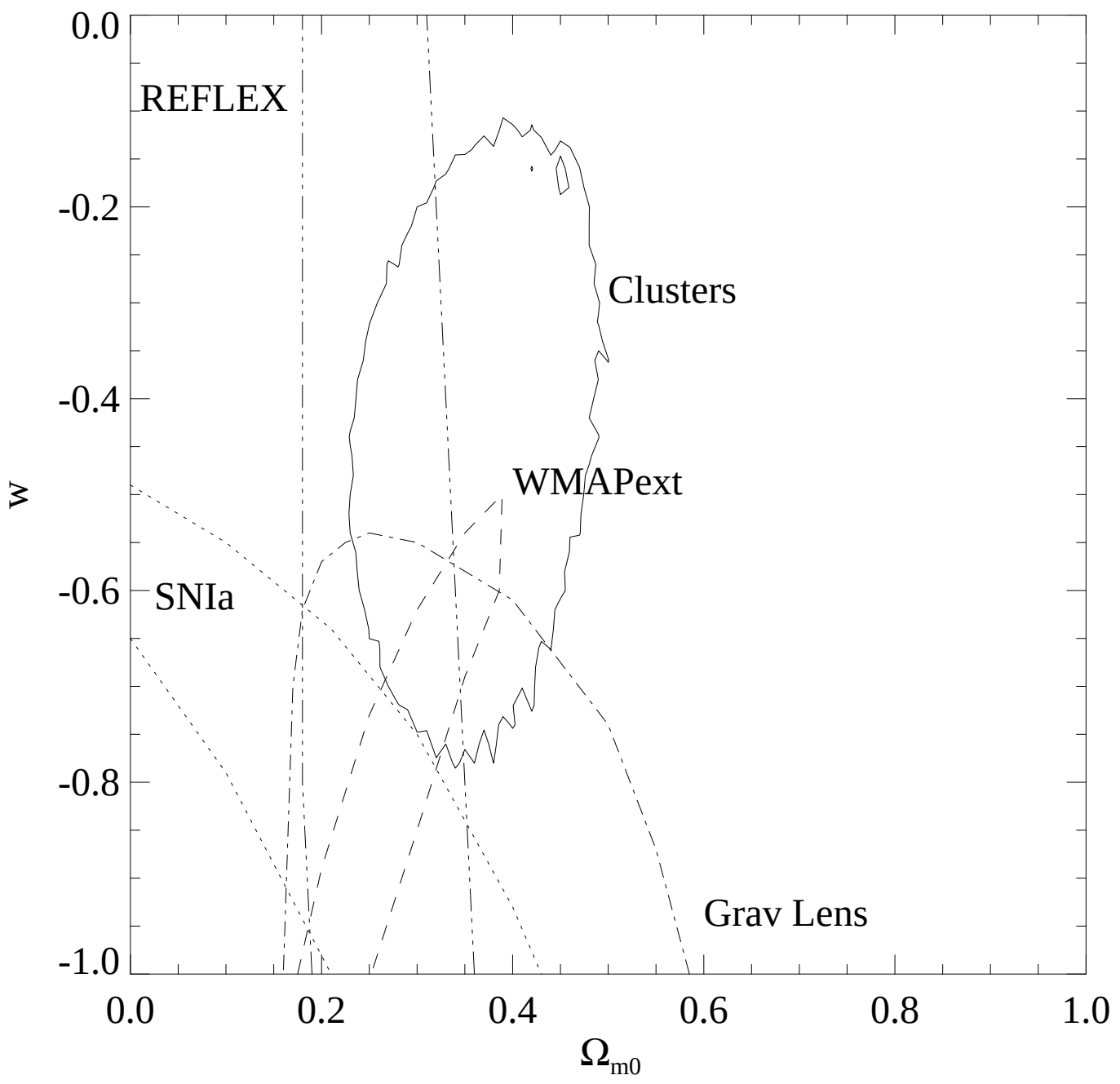

Figure 6. Constraints on w and $\Omega_{m0}$ are shown as 1 $\sigma$ contours from cluster temperature evolution (solid; Fig. 5), REFLEX (local) clusters (double dot dashed; Schuecker et al., 2003b), supernovae (dotted; Turner & Riess, 2002), gravitational lensing of radio sources (dot dashed; Chae et al., 2002) and cosmic microwave background (dashed; Spergel et al., 2002). All results assume $\Omega_{Q0} + \Omega_{m0} = 1$.

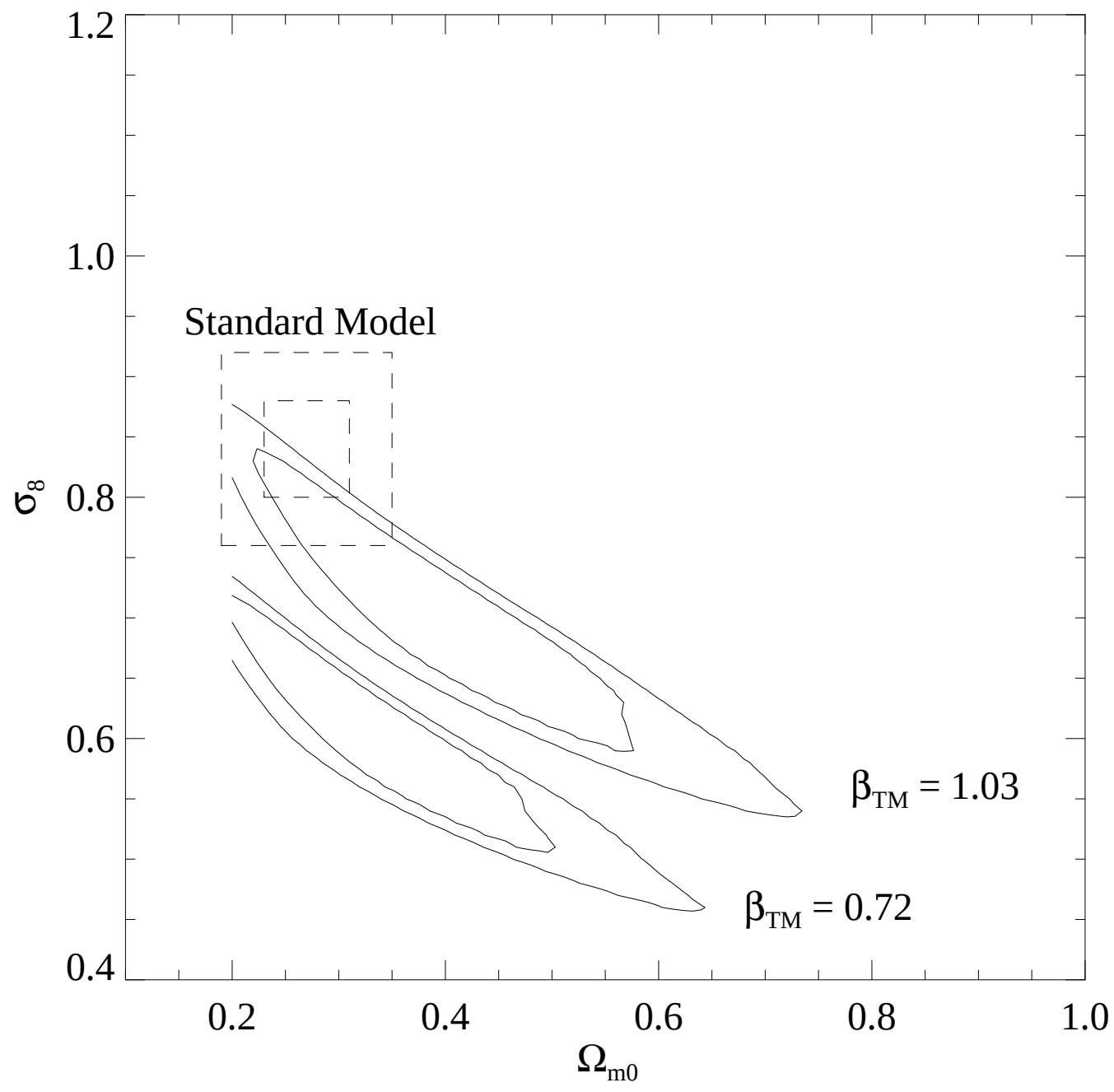

Figure 7. Constraints on $\sigma_8$ and $\Omega_{m0}$ are shown as 1 and 2 $\sigma$ contours for three interesting parameters (i.e. $\Delta S$ = 3.5, 8.0) from cluster temperature evolution for two values of $\beta_{TM}$ (solid). The 1 and 2 $\sigma$ contours for the standard model values are also shown (dashed).

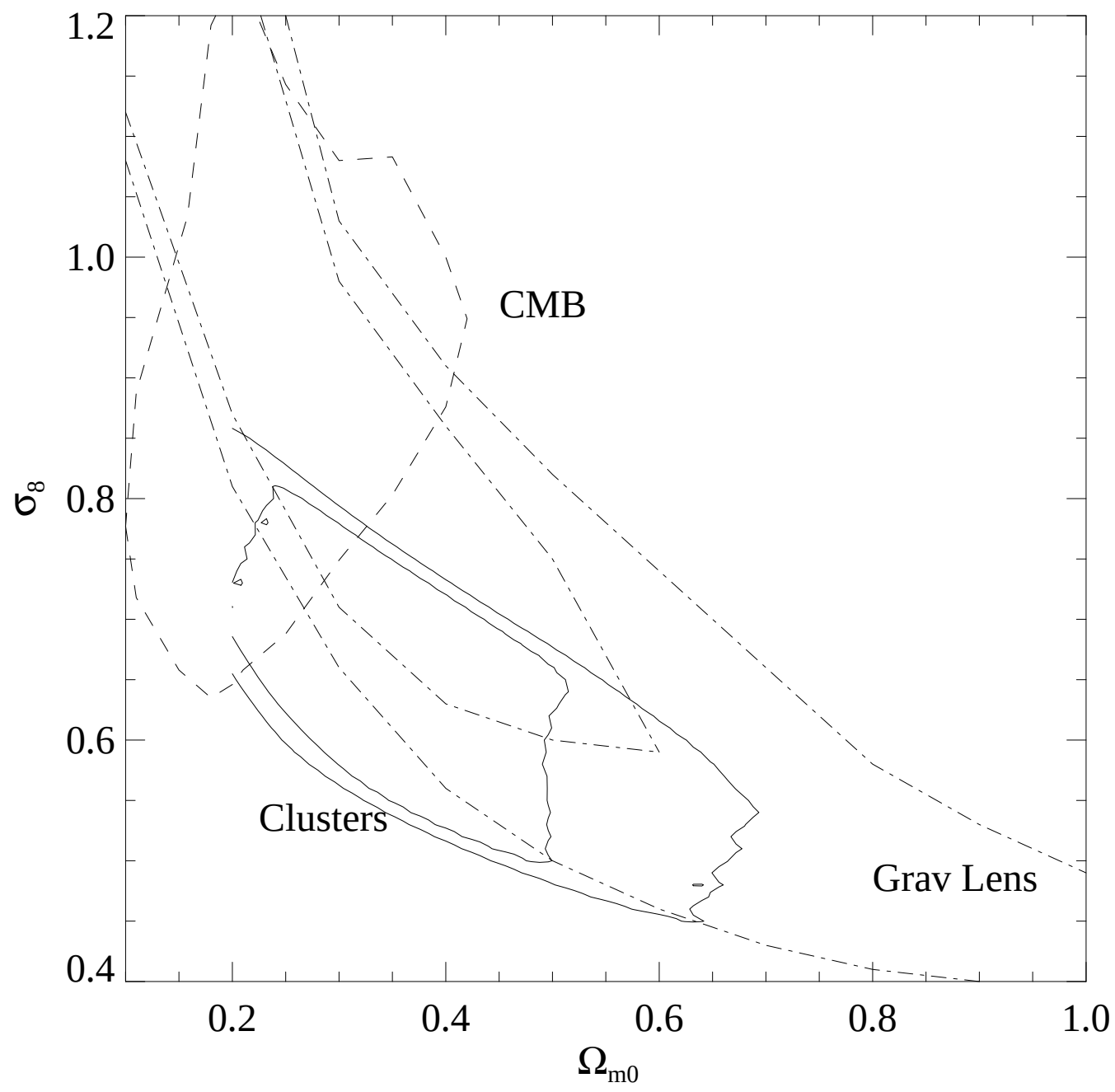

Figure 8. Constraints on $\sigma_8$ and $\Omega_{m0}$ are shown as 1 and 2 $\sigma$ contours from cluster temperature evolution (solid; marginalized over $\beta_{TM}$) and gravitational lensing (dot dashed; Hoekstra, Yee & Gladders, 2002) and a 2 $\sigma$ contour from the cosmic microwave background (dashed; Bridle et al., 2003, no 1 $\sigma$ constraint quoted). The cluster constraints are for 3 interesting parameters (i.e. $\Delta S$ = 3.5, 8.0).

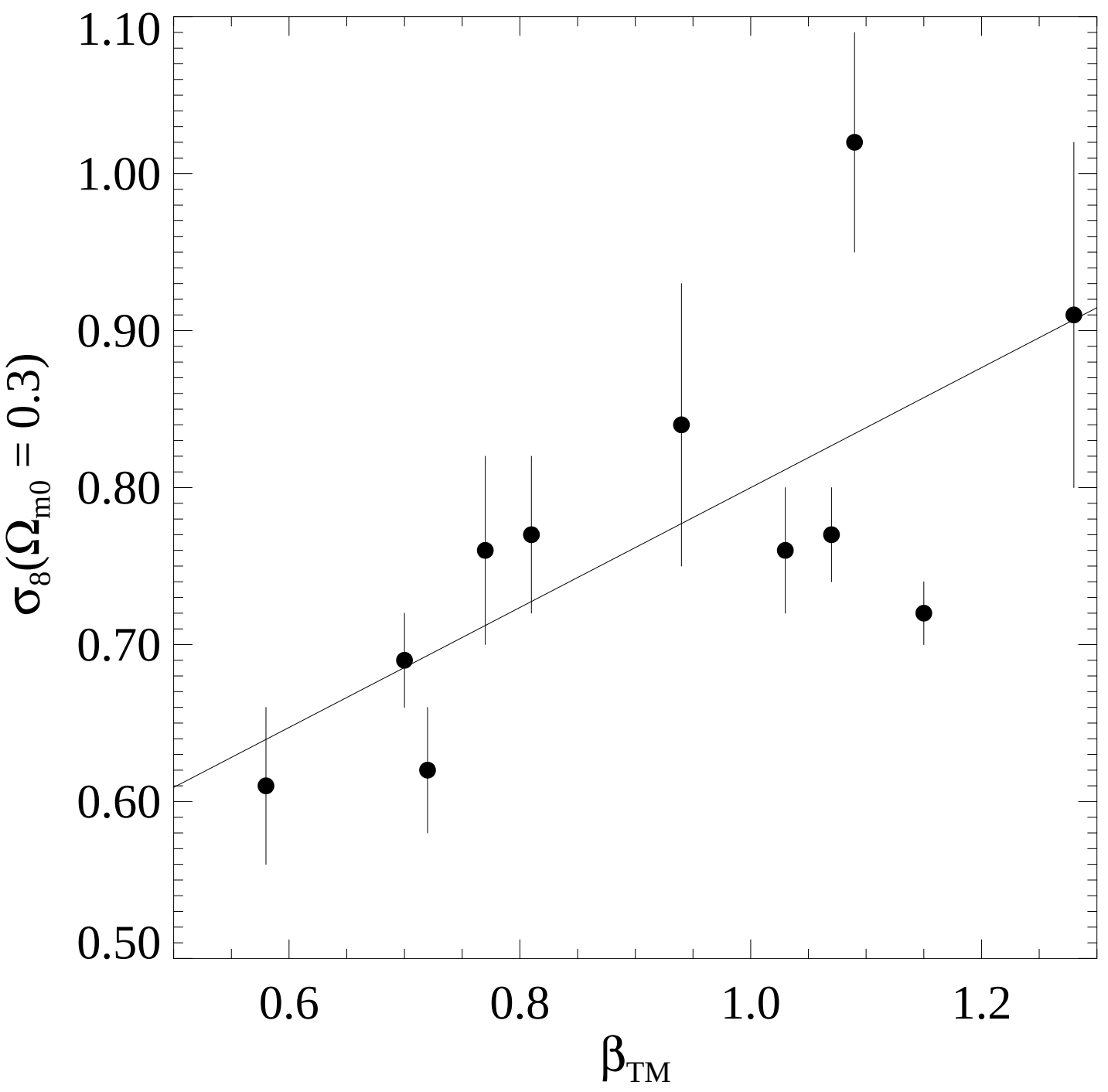

Figure 9. Correlation of $\sigma_8$ (for $\Omega_{m0} = 0.3$) against the normalization of the mass – temperature relation assumed when determining it. The line is the best linear fit to the unweighted data, $\sigma_8 = 0.42 + 0.38\beta_{TM}$. Most of the scatter among the reported measurements of $\sigma_8$ is simply due to the assumed mass – temperature relation normalization.